\documentclass{article}

\usepackage{graphics}

\catcode`\@=11
\ifcase \@ptsize
\or 
\or 
\fi
\advance\textheight by \topskip

\catcode`\@=12
\begin{document}
\title{\Large \bf The Superfluid State of a Bose Liquid
as a Superposition of a Suppressed Bose-Eistein Condensate
and an Intensive Pair Coherent Condensate}
\author {\large E.A.~Pashitskii$^1$, S.V.~Mashkevich$^2$,
S.I.~Vilchynskyy$^3$\\[0.5cm]
$^1$\it Institute of Physics, NAS of Ukraine, Kiev 03022, Ukraine \\
\it pashitsk@iop.kiev.ua\\
$^2$\it N.N.Bogolyubov Institute of Theoretical Physics, Kiev 03143,
Ukraine\footnote{Present address:
Schr\"odinger, 120 West 45th St., New York, NY 10036, USA.} \\
\it mash@mashke.org\\
$^3$\it Taras Shevchenko National Kiev University, Kiev 03022, Ukraine \\
\it sivil@phys.univ.kiev.ua }
\maketitle
\date{}
\begin{abstract}
A self-consistent model of  the superfluid (SF) state of a Bose liquid
with strong interaction between bosons is considered,
in which at $T=0$, along with a
weak single-particle Bose-Einstein condensate (BEC), there exists
an intensive pair coherent condensate (PCC) of bosons, analogous to the
Cooper pair condensate of fermions. Such a PCC emerges due to an effective
attraction between bosons in some regions of momentum space, which
results from an oscillating sign-changing momentum dependence of the
Fourier component $V(p)$ of the interaction potentials $U(r)$ with
the inflection points in the radial dependence.
The collective many-body effects of renormalization (``screening'') of the
initial interaction, which are described by the bosonic polarization
operator $\Pi(\mathbf{p},\omega)$,
lead to a suppression of the repulsion [$V(p)>0$] and an
enhancement of the effective attraction  [$V(p)<0$]
in the respective domains of nonzero momentum transfer,
due to the negative sign of the real part
of $\Pi(\mathbf{p},\omega)$ on the ``mass shell'' $\omega=E(p)$.
The ratio of the BEC density $n_0$
to the total particle density $n$ of the Bose liquid
is used as a small parameter of the model, $n_0/n\ll 1$, unlike in the
Bogolyubov theory of a quasi-ideal Bose gas, in which the small
parameter is the ratio of the number of supracondensate excitations
to the number of particles in an intensive BEC, $(n-n_0)/n_0\ll 1$.
A closed system of nonlinear integral equations for the normal
$\tilde\Sigma_{11}(\mathbf{p}, \omega)$ and anomalous
$\tilde\Sigma_{12}(\mathbf{p}, \omega)$ self-energy parts is obtained
with account for the terms of first order in the BEC density.
A renormalized perturbation theory is used, which is built on
combined hydrodynamic (at  $\mathbf{p}\to 0$) and field (at
$\mathbf{p}\ne 0 $) variables with
analytic functions $\tilde\Sigma_{ij}(\mathbf{p}, \epsilon)$
at $\mathbf{p}\to 0 $ and $\epsilon\to 0$ and a nonzero SF order
parameter $\tilde\Sigma_{12}(0,0)\ne 0$, 
proportional to the density $\rho_s$ of the SF component
which is a superposition of the BEC and PCC.
In the framework of the ``soft spheres'' model with the single fitting
parameter---the value of the repulsion potential at $r=0$, a
theoretical quasiparticle spectrum $E(p)$
is obtained, which is in good accordance with the experimental spectrum
$E_\mathrm{exp}(p)$ of elementary excitations in superfluid $^4$He.
\end{abstract}


\newpage
\large

\section{Introduction}
An ab initio computation of
the spectrum of elementary excitations in the
superfluid (SF) $^4$He Bose liquid
remains an actual problem nowadays, despite certain recent
successes in that direction, like an
excellent agreement with experimental data
in the region of the roton minimum obtained by the Monte Carlo method
making use of the so-called ``shadow wave function'' \cite{1}
and by the corellation basic function method \cite{2} employing
modern interatomic potentials for $^4$He \cite{3}--\cite{5}.
At the same time, a microscopic field perturbation
theory \cite{6}-\cite{8} calculation of
the long-wave phonon part of the spectrum $E(p)\simeq c_1 p$, where $c_1$
is the speed of first (hydrodynamic) sound in liquid $^4$He,
faces principal difficulties. This is due to the fact that
nonrenormalized perturbation theory gives rise to infrared
divergencies and nonanalyticities at
$p\to 0$ and $\epsilon\to 0$ \cite{9}--\cite{12}, which can be
cured with the technique of ``combined variables'' \cite{13}.
In the long-wave limit $(p\to 0)$, those variables
reduce to the hydrodynamic variables of macroscopic quantum
hydrodynamics \cite{14}, while in the short-wave domain they correspond
to the bosonic quasiparticle creation and annihilation operators.

On the other hand, according to numerous precise
experimental data on neutron inelastic scattering
\cite{15}--\cite{18} and to results in quantum evaporation
of $^4$He atoms \cite{19}, the maximal density $\rho_0$ of the
single-particle Bose-Einstein condensate (BEC) in the $^4$He
Bose liquid even at very low temperatures $T\ll T_\lambda$
does not exceed $10 \%$ of the total density $\rho$ of liquid
$^4$He, whereas the density of the SF component
$\rho_s \to \rho $ at $T\to 0$ \cite{20}.
Such a low density of the BEC is implied by strong interaction
between $^4$He atoms and is an indication of the fact that
the quantum structure of the part of the SF condensate in He~II
carrying the ``excess'' density
$(\rho_s-\rho_0)\gg \rho_0$ calls for a more thorough investigation.
The questions discussed in this paper are both those of the
quantum structure of the SF state in  a Bose liquid at $T=0$
and the self-consistent
calculation of the spectrum $E(p)$ of elementary excitations
in the framework of renormalized field perturbation theory
\cite{9}--\cite{13}.

Our approach is based on the
microscopic model developed in Refs.~\cite{21}--\cite{22},  of
superfluidity of a Bose liquid with a suppressed BEC and an
intensive pair coherent condensate (PCC), which can arise from
a sufficiently strong effective attraction between bosons in some
domains of momentum space (see below) and is analogous to
the Cooper condensate in a Fermi liquid with attraction between
fermions near the Fermi surface \cite{23}. As a small parameter,
one uses the ratio of the BEC density to the total Bose liquid
density $(n_0 / n)\ll 1$, unlike in the Bogolyubov theory \cite{24}
for a quasi-ideal Bose gas, in which the small parameter is
the ratio of the number of supracondensate excitations to the
 density of the intensive BEC, $(n-n_0)/n_0\ll 1$.
Because of this, the SF state within the model at hand can be
described by a ``short'' self-consistent system of Dyson-Belyaev
equations for the normal and anomalous Green functions
$\tilde G_{ik}$ and self-energy parts
$\tilde\Sigma_{ij} (\mathbf{p},\omega)$ without account for the diagrams
of second and higher orders in the BEC density.
In this case, the SF component $\rho_s$ is a superposition of the
``weak'' single-particle BEC and an intensive ``Cooperlike'' PCC
with coinciding phases (signs) of the corresponding order parameters.
The pair interaction between bosons was chosen in the form of a
finite repulsive potential in the ``semitransparent'', or ``soft''
spheres model, whose Fourier component $V(p)$ is an
oscillating sign-changing function of momentum transfer $p$ due
to mutual quantum diffraction of particles.

As a result of renormalization (``screening'') of the initial
interaction $V(p)$ due to multiparticle collective correlations,
which are described by the boson polarization operator
$\Pi (\mathbf{p}, \omega)$,
the interaction gets suppressed in
the domains of momentum space
where $V(p)>0$, and enhanced where $V(p)<0$. Such a suppression of
repulsion and enhancement of attraction is implied by the negative
sign of the real part of $\Pi (\mathbf{p}, \omega)$ on the ``mass shell''
$\omega=E(p)$ for a decayless quasiparticle spectrum.
It is shown that the integral contribution of the domains of
effective attraction in the renormalized sign-changing interaction
$$\tilde V(\mathbf{p})=V(p)\left[1-V(p)
\Re\,\Pi (\mathbf{p}, E(p))\right]^{-1}$$
can be sufficient for the formation of an intensive bosonic PCC in
momentum space (although not for the formation of bound boson pairs
in real space).

Self-consistent numerical calculations of the boson self-energy,
polarization operator, pair order parameter, and quasiparticle
spectrum at $T=0$, involving an iteration scheme with the single
fitting parameter---the value of the repulsion potential
at $r=0$, have allowed us
to find conditions for the theoretical spectrum $E(p)$ to coincide
with the experimentally observed elementary excitation spectrum
in $^4$He \cite{30}--\cite{Pearce}. The roton
minimum in the quasiparticle spectrum $E(p)$, which corresponds to 
a maximum in the structural form factor $S(q)$ of a Bose liquid,
turns out to be directly associated with the first negative minimum of the 
Fourier component of the renormalized potential $\tilde V(\mathbf{p})$
of pair interaction between bosons.

\section{Equations for the Green functions and self-energy parts
in a Bose liquid with a suppressed BEC and an intensive PCC in the
renormalized perturbation theory}

The main difficulty of the microscopic description of the SF state
of a Bose liquid with a nonzero BEC is the fact that applying
perturbation theory directly \cite{6} leads, as was shown in
Refs.~\cite{9}--\cite{12}, to divergences and non-analyticities
at small energies $\epsilon \to 0$ and momenta $\mathbf{p} \to 0$ and, as
a consequence, to erroneous results in the calculations of different
physical quantities.
	Thus, for example, for a Bose system with weak interaction, when the ratio of
the mean potential energy $V(p_0)p_0^3$ ($p_0$ being a
typical momentum transfer) to the corresponding kinetic energy
$p_0^2/2m$ of the bosons is small, the zeroth-approximation polarization
operator $\Pi (\mathbf{p}, \omega)$ and the density-density
response function $\tilde \Pi (\mathbf{p},\omega)$ calculated to
the first order in the small parameter of interaction
$\xi =m p_0V(p_0)\ll1$,
are logarithmically divergent at $p\to 0$,
 $ \omega\to 0$, whereas
the exact values $\Pi(0,0)$ and $\tilde \Pi(0, 0)$ are finite \cite{10}:
\begin{equation}
\Pi(0, 0)=-\frac{\partial n}{\partial \mu}=-\frac{n}{mc^2}\;; \qquad
\tilde\Pi(0, 0)=\frac{n}{m(c_B^2-c^2)}\;,
\label{1}
\end{equation}
where $n$ is the total concentration of bosons, $\mu$
the chemical
potential, $c_B=\sqrt{nV_0/m}$ the velocity of sound in
 the Bogolyubov
approximation for a weakly nonideal Bose gas \cite{24},
$V_0\equiv V(0)$ the zero Fourier component of the potential,
and $c$ the speed of sound in the $\mathbf{p}\to 0$ limit for the
spectrum of elementary excitations $\epsilon (p)\simeq c|\mathbf{p}|$
in the Belyaev theory \cite{6}:
\begin{equation}
c=\sqrt{\Sigma_{12}(0,0)/m^*}\;.
\label{2}
\end{equation}
Here $m^*$ is the effective mass of quasiparticles, which is
determined by the relation \cite{7}
\begin{equation}
\frac{1}{m^*}=\frac{2}{B}\left[
\frac{1}{2m}+\frac{\partial \Sigma_{11}(0,0)}{\partial |\mathbf{k}|^2 }
-\frac{\partial \Sigma_{12}(0,0)}{\partial |\mathbf{k}|^2 }
\right]\;,
\label{3}
\end{equation}
where $\Sigma_{11}(0,0)$ is
the normal  self-energy part (at $k\to 0$, $\epsilon \to 0$),
and
\begin{equation}
B=\left[1-\frac{\partial \Sigma_{11}(0,0)}{\partial\epsilon }\right]^2
-\Sigma_{11}(0,0)\frac{\partial^2 \Sigma_{12}(0,0)}{\partial\epsilon^2}+
\frac{1}{2}\frac{\partial^2 }{\partial \epsilon^2}\left[
\Sigma_{12}(0,0)\right]^2\;.
\label{4}
\end{equation}
The model, considered in Ref.~\cite{6}, of a dilute Bose system
of hard spheres with a small parameter
$\beta=\sqrt{n/k_0^3}\ll1$, in which there is
a possibility to exclude the infinite repulsion
by means of a summation of the ``ladder'' diagrams,
leads to a finite
value of $\Sigma_{12}(0,0)$ in the zeroth approximation in $\beta$:
\begin{equation}
\Sigma_{12}(0,0)= \frac{4\pi a_0}{m}n_0\;,
\label{5}
\end{equation}
$a_0$ being the vacuum scattering amplitude
and $n_0$   the concentration of bosons in the BEC ($\rho_0=mn_0$).

At the same time, in Ref.~\cite{25}, taking into account an exact
thermodynamic equation
\begin{equation}
\frac{\partial \Sigma_{11}(0,0)}{\partial \epsilon}=
-\left(
\frac{\partial n_1}{\partial n_0}\right)_{\mu}=
1-\frac{1}{n_0}\Sigma_{12}(0,0)\frac{d n_0}{d\mu}\;,
\label{6}
\end{equation}
where $n_1=n-n_0$ is the concentration of supracondensate bosons,
exact asymptotic relations were obtained for the normal and anomalous
single-particle Green functions:
\begin{equation}
G_{11}(\mathbf{p}\to 0)=-G_{12}(\mathbf{p}\to 0)=
\frac{n_0 mc^2}{n(\epsilon^2-c^2\mathbf{p}^2+i\delta)}\;;
\quad c^2=\frac{n}{m}\frac{d\mu}{dn}\;.
\label{7}
\end{equation}

However, it was shown in Refs.~\cite{9}--\cite{11}
that at $\mathbf{p}=0$, $\epsilon=0$ the anomalous self-energy
part is precisely equal to zero, $\Sigma_{12}(0,0)\equiv 0$.
Problems then emerge with the determination
of the velocity of sound (\ref{2}) and  the
asymptotic formulas for  $G_{11}(\mathbf{p},\epsilon)$
and $G_{12}(\mathbf{p},\epsilon)$  at $(\mathbf{p},\epsilon)\to 0$
\cite{7}:
\begin{equation}
G_{11}(\mathbf{p}\to 0)=-G_{12}(\mathbf{p}\to
0)=\frac{\Sigma_{12}(0,0)}{B(\epsilon^2-c^2\mathbf{p}^2+i\delta)}\;,
\label{8}
\end{equation}
because at $\Sigma_{12}(0,0)=0$, relations (\ref{4}) and (\ref{6})
reduce to identities
\begin{equation}
\frac{\partial \Sigma_{11}(0,0)}{\partial \epsilon}\equiv 1\;,
\qquad B\equiv 0\;,
\label{9}
\end{equation}
so that Eqs.~(\ref{2}) and (\ref{8}) with account for (\ref{3})
contain uncertainties of the $0/0$ type.

With the purpose of fixing these controversies, as well as the
infrared divergences of $\Pi(\mathbf{p},\epsilon)$ and nonanalyticities in
$\Sigma_{ij}(\mathbf{p},\epsilon)$ at $(\mathbf{p},\epsilon)\to 0$
emerging in the nonrenormalized theory, a renormalization procedure for the
field perturbation theory was worked out in Ref.~\cite{12},
employing the method of ``combined variables'' \cite{13}.
The perturbation theory built on such ``adequate'' field variables
does not suffer from infrared divergences at $(\mathbf{p},\epsilon)\to 0$,
whose source at $T=0$ is the divergence of long-wave quantum fluctuations
(acoustic Goldstone oscillations).
Such oscillations are associated with a spontaneous breakdown
of continuous gauge and translational symmetries in the SF state
of a Bose system with a uniform coherent condensate and corresponded to
the hydrodynamic first sound in liquid $^4$He,
propagating with the velocity of
$c_1\simeq 236 \mbox{ m/s}$.
The choice of combined variables \cite{13} leads to the
renormalized anomalous self-energy part
$\tilde \Sigma_{12}(\mathbf{p},\epsilon)$ which
does not vanish at $(\mathbf{p},\epsilon)=0$.
Then one can formally restore all the results
of the nonrenormalized field theory \cite{6}--\cite{7}, but now in
terms of the renormalized quantities $\tilde G_{ik}(\mathbf{p},\epsilon)$
and $\tilde \Sigma_{ik}(\mathbf{p},\epsilon)$,
which do not contain singularities at
$(\mathbf{p},\epsilon)\to 0$ (save for the pole part
$\tilde G_{ik}(\mathbf{p},\epsilon) \sim\vert  p\vert ^{-2}$).
In particular, the squared velocity of first sound $c_1$
at $T \to 0$ must be equal to
\begin{equation}
c_1^2=
\frac{\tilde \Sigma_{12}(0,0)}{\tilde m^*}\;,
\label{10}
\end{equation}
where the renormalized effective mass $\tilde m^*$ is determined
by relations (\ref{3}) and (\ref{4}) with
$\tilde \Sigma_{ik}(0,0)$ substituted for $ \Sigma_{ik}(0,0)$
in Eq.~(\ref{2}).
In view of the aforesaid, we will work with the
combined variables \cite{13},
\begin{equation}
\tilde \Psi(x)= \tilde \Psi_\mathrm{L} (x)+\tilde \Psi_\mathrm{sh} (x)\;,
\label{11}
\end{equation}
where
\begin{equation}
\begin{array}{c}
\displaystyle
\tilde \Psi_\mathrm{L} (x)= \sqrt{\left\langle
\tilde n_\mathrm{L}\right\rangle}\left[
1+\frac{\tilde n_\mathrm{L}-\left\langle \tilde n_\mathrm{L}\right\rangle}
{2\left\langle \tilde n_\mathrm{L}\right\rangle} +i\tilde \phi_\mathrm{L}
\right]; \quad \tilde \Psi_\mathrm{sh}=
\psi_\mathrm{sh}e^{-i\tilde\phi_\mathrm{L}}\;;\\[12pt]
\displaystyle
\psi_\mathrm{sh}=\psi-\psi_\mathrm{L};\quad
\psi_\mathrm{L}(\mathbf{r})=\frac{1}{\sqrt{V}}
\sum_{|\mathbf{k}|<k_0} a_{\mathbf{k}} e^{i\mathbf{k} \mathbf{r}}=
\sqrt{\left\langle \tilde n_\mathrm{L}\right\rangle}
e^{i\tilde \phi_\mathrm{L}}\;.
\end{array}
\label{12}
\end{equation}
Such an approach means that the separation of the Bose system into a
macroscopic coherent condensate and a gas of supracondensate excitations
is made not on the statistical level,
like in the case of a weakly nonideal Bose
gas \cite{24}, but on the level of ab initio field operators,
which are used to construct a microscopic theory of the
Bose liquid.

The system of Dyson-Belyaev equations \cite{6}--\cite{7}, which allows
one to express the normal $\tilde G_{11}$ and anomalous $\tilde G_{12}$
renormalized single-particle boson Green functions in terms of the
respective self-energy parts $\tilde \Sigma_{11}$ and $\tilde \Sigma_{12}$,
has the form (Fig.~1):
\begin{equation}
\tilde G_{11} (\mathbf{p},\epsilon)=\left[G_0^{-1}(-\mathbf{p},-\epsilon)-
\tilde \Sigma_{11}(-\mathbf{p},-\epsilon)\right]/Z(\mathbf{p},\epsilon)\;;
\label{13}
\end{equation}
\begin{equation}
\tilde G_{12}(\mathbf{p},\epsilon)=
\tilde \Sigma_{12}(\mathbf{p},\epsilon)/Z(\mathbf{p},\epsilon)\;.
\label{14}
\end{equation}
Here
\begin{equation}
Z(\mathbf{p},\epsilon)=
\left[G_0^{-1}(-\mathbf{p},-\epsilon)-
\tilde \Sigma_{11}(-\mathbf{p},-\epsilon)\right]
\left[G_0^{-1}(\mathbf{p},\epsilon)-
\tilde \Sigma_{11}(\mathbf{p},\epsilon)\right]
-\vert \tilde \Sigma_{12}(\mathbf{p},\epsilon) \vert ^2\;;
\label{15}
\end{equation}
\begin{equation}
G_0^{-1}(\mathbf{p},\epsilon)=
\left[\epsilon-\frac{\mathbf{p}^2}{2m}+\mu+i\delta \right]\qquad
(\delta\to +0)\;,
\label{16}
\end{equation}
where $\mu$ is the chemical potential of the quasiparticles,
which satisfies the Hugengoltz-Pines relation \cite{26}:
\begin{equation}
\mu=\tilde\Sigma_{11}(0,0)-\tilde\Sigma_{12}(0,0)\;.
\label{17}
\end{equation}
Due to a strong hybridization of the single-particle and collective
branches of elementary excitations in the Bose liquid with a
finite BEC $ (n_0\ne 0)$, the poles of the two-particle and all
multiparticle Green functions
coincide with the poles of the single-particle Green functions
$ \tilde G_{ik}(\mathbf{p}, \epsilon)$
\cite{7}--\cite{8}.
Therefore the spectrum of all elementary excitations with zero
spirality is determined by the zeros of the function
$Z(\mathbf{p}, \epsilon )$:
\begin{equation}
E(p)=\left\{ \left[\frac{\mathbf{p}^2}{2m}+\tilde \Sigma_{11}^s
(\mathbf{p}, E(p)) -\mu \right]^2- \vert \tilde \Sigma _{12}
(\mathbf{p},  E(p)) \vert ^2\right\}^{1/2}
+\tilde \Sigma_{11}^a (\mathbf{p}, E(p))\;,
\label{18}
\end{equation}
where
$$
\tilde \Sigma_{11}^{s,a} (\mathbf{p}, \epsilon) =\frac{1}{2}
\left[\tilde \Sigma_{11}(\mathbf{p}, \epsilon) \pm
\tilde \Sigma_{11}(-\mathbf{p}, -\epsilon) \right]\;.$$
The $(+)$ and $(-)$ signs correspond to the symmetric
$\tilde \Sigma_{11}^s$ and antisymmetric
$\tilde \Sigma_{11}^a$ parts
of $\tilde \Sigma_{11}$, respectively.
Relation (\ref{17}) ensures the acoustic dispersion law for the
quasiparticles spectrum (\ref{18}) at $\mathbf{p}\to 0$
with the sound velocity (\ref{10}).
As was shown in Ref.~\cite{21}, for a Bose liquid with strong enough
interaction between particles, when the BEC is strongly suppressed,
one can, when defining $\tilde \Sigma_{ik}(\mathbf{p}, \epsilon)$ in the
form of a sequence of irreducible diagrams containing condensate
lines, restrict oneself, with good precision, to the
first (lowest) terms in the expansion over the small BEC density
($n_0\ll n$). Such an approximation is exactly opposite to the
Bogolyubov approximation \cite{24} for a weakly nonideal Bose gas with
an intensive BEC, when $n_0\simeq n$.
As a result, up to terms of first order in the small parameter
$n_0/n\ll 1$, for a Bose liquid one gets the ``trimmed''
system of equations for $\tilde{\Sigma}_{ik}$ \cite{21}, \cite{22}
(see Fig.2):
\begin{equation}
\tilde \Sigma_{11}(\mathbf{p}, \epsilon) =n_0\Lambda (\mathbf{p}, \epsilon)
 \tilde V(\mathbf{p}, \epsilon)+
 n_1 V(0)+\tilde \Psi _{11}(\mathbf{p}, \epsilon)\;;
\label{19}
\end{equation}
\begin{equation}
\tilde \Sigma_{12}(\mathbf{p}, \epsilon) =
n_0\Lambda (\mathbf{p}, \epsilon) \tilde V(\mathbf{p}, \epsilon)+
\tilde \Psi_{12} (\mathbf{p}, \epsilon)\;,
\label{20}
\end{equation}
where
\begin{equation}
\tilde{\Psi}_{ij}(\mathbf{p}, \epsilon)=
i\int \frac{d^3 \mathbf{k}}{(2\pi)^3}\int \frac{d\omega}{2\pi}\,
G_{ij}(\mathbf{k})\tilde V (\mathbf{p}-\mathbf{k}, \epsilon-\omega)
\Gamma(\mathbf{p},\epsilon, \mathbf{k}, \omega )\;,
\label{21}
\end{equation}

\begin{equation}
\tilde V(\mathbf{p}, \epsilon) =V(p)\left[1- V(p)
 \Pi(\mathbf{p}, \epsilon)\right]^{-1}\;.
\label{22}
\end{equation}
Here $V(p)$ is the Fourier component
of the input potential of pair interaction of bosons,
$\tilde V(\mathbf{p}, \epsilon)$ is the renormalized
(``screened''), due to multiparticle collective effects,
Fourier component of the retarded (nonlocal) interaction;
$\Pi (\mathbf{p}, \epsilon)$ is the boson polarization operator:
\begin{equation}
\begin{array}{c}
\displaystyle
\Pi(\mathbf{p}, \epsilon)=
i\int \frac{d^3 \mathbf{k}}{(2\pi)^3}\int \frac{d\omega}{2\pi}\,
\Gamma(\mathbf{p}, \epsilon,\mathbf{k},\omega)
\\[12pt]
{}\times\left\{G_{11}(\mathbf{k}, \omega)
G_{11}(\mathbf{k}+\mathbf{p}, \epsilon+\omega)
+G_{12}(\mathbf{k}, \omega)
G_{12}(\mathbf{k}+\mathbf{p}, \epsilon+\omega)\right\}\;;
\end{array}
\label{23}
\end{equation}
$\Gamma (\mathbf{p},\, \epsilon ;\,\mathbf{k},\, \omega)$ is the vertex
part, which describes multiparticle correlations;
$\Lambda (\mathbf{p}, \epsilon)=\Gamma (\mathbf{p}, \epsilon, 0, 0)=
\Gamma (0,0,\mathbf{p}, \epsilon)$ ,\,
and $n_1$ is the number of supracondensate particles
($n_1\gg n_0$), which is determined from the condition of
conservation of the total number of particles:
\begin{equation}
n=n_0+n_1=n_0+ i\int\frac{d^3 \mathbf{k}}{(2\pi)^3}
\int \frac{d\omega}{2\pi}\,G_{11}(\mathbf{k}, \omega)\;.
\label{24}
\end{equation}
In the sequel, as well as in Ref.~\cite{21},
in the integral relations (\ref{21})
and (\ref{23}) we will only consider the residues at
poles of single-particle Green functions
$\tilde G_{ij}(\mathbf{p}, \epsilon)$,
neglecting the contributions of eventual poles of the functions
$\Gamma (\mathbf{p},\epsilon, \mathbf{k}, \omega )$
and $\tilde V(\mathbf{p}, \epsilon)$,
which do not coincide with those of
$\tilde G_{ij}(\mathbf{p}, \epsilon )$. As a result,
taking into account relations (\ref{13})--(\ref{20}),
~Eqs.~(\ref{21}) on the mass shell
$\epsilon=E(p)$ assume the following form (at $T=0$):
\begin{eqnarray}
\tilde{\Psi}_{11}(\mathbf{p}, E(p))&=&
\frac{1}{2}\int \frac{d^3 \mathbf{k}}{(2\pi)^3}\,
\Gamma (\mathbf{p}, E(p);\mathbf{k}, E(k))
\nonumber \\
&& {} \times \tilde{V}(\mathbf{p}- \mathbf{k},
E (\mathbf{p})- E(k)) \left[\frac{A(\mathbf{k}, E(k))}
{E(k)}-1\right]\;;
\label{25}
\end{eqnarray}
\begin{eqnarray}
\tilde{\Psi}_{12}(\mathbf{p},  E(p))&=&
 -\frac{1}{2}\int \frac{d^3  \mathbf{k}}
{(2\pi)^3} \Gamma (\mathbf{p}, E(p);
\mathbf{k}, E(k)) \tilde{V}(\mathbf{p}- \mathbf{k},
E(p)- E (\mathbf{k}))
\nonumber \\
&& {}\times \frac{n_0\Lambda (\mathbf{k}, E(k))
\tilde{V}(\mathbf{k}, E(k))+
\tilde{\Psi}_{12}(\mathbf{k},E(k))}{E(k)}\;,
\label{26}
\end{eqnarray}
where
\begin{eqnarray}
E(p)&=&\left\{
A^2(\mathbf{p}, E(p))-\left[
n_0\Lambda (\mathbf{p}, E(p)) \tilde{V}(\mathbf{p}, E(p))+
\tilde\Psi_{12}(\mathbf{p}, E(p))\right]^2\right\}^{1/2}
\nonumber \\
&&{}+\frac{1}{2}\left[
\tilde\Psi_{11}(\mathbf{p},E(p))-
\tilde\Psi_{11}(-\mathbf{p},-E(p))
\right];
\label{27}
\end{eqnarray}
\begin{eqnarray}
A(\mathbf{p},E(p))&=&n_0\Lambda (\mathbf{p}, E(p))\tilde{V}
(\mathbf{p}, E(p))
\nonumber \\
&& {}+\frac{1}{2}
\left[\tilde{\Psi}_{11}(\mathbf{p}, E(p))+
\tilde{\Psi}_{11}(-\mathbf{p}, -E(p))
\right] -\tilde{\Psi}_{11}(0,0)
+\tilde{\Psi}_{12}(0,0)+\frac{\mathbf{p}^2}{2m}\;.
\nonumber \\
\label{28}
\end{eqnarray}
The total quasiparticle concentration in the Bose liquid is
determined by the relation
\begin{equation}
n=n_0+\frac{1}{2} \int \frac{d^3 \mathbf{k}}{(2\pi)^3}
\left[\frac{A(\mathbf{k},E(k))}{E(k)}-1\right]\;.
\label{29}
\end{equation}
{}From Eqs.~(\ref{27}) and (\ref{28}) it follows that the quasiparticle
spectrum $E(p)$, because of the analyticity of the functions
$\tilde \Psi_{ij}(\mathbf{p}, \epsilon)$, is acoustic at $p\to 0$,
and its structure at $p\ne 0$ depends essentially on the
features of the renormalized pair interaction of bosons.
The theoretical spectrum (\ref{27}) must be close to the experimental
spectrum of elementary excitations in the $^4$He Bose liquid
\cite{30}--\cite{Pearce} if this model is to be
applicable for the description of the SF state in $^4$He (see below).

Note that when the BEC is totally absent ($n_0=0$),
equation (\ref{26}) becomes homogeneous and degenerate with
respect to the phase of $\tilde\Psi_{12}(p)$. It is then
akin to the Bethe-Goldstone integral equation for a pair of
particles in momentum space
\begin{equation}
\Psi(\mathbf{p})=
 -\int \frac{d^3  \mathbf{k}}
{(2\pi)^3}  V(\mathbf{p}- \mathbf{k})
 \frac{\Psi(\mathbf{k})}{2 E(k)-\Omega}\;,
\end{equation}
with zero binding energy, $\Omega=0$,
which has a nontrivial solution $\Psi(\mathbf{p})\ne 0$
only when there is attraction $V(\mathbf{q})<0$ in a broad
enough region of momentum transfer $\mathbf{q}$.
By virtue of this analogy, the function $\tilde\Psi_{12}(p)$
at $n_0=0$ can be taken to be the PCC order parameter
\cite{21}--\cite{22},
which describes condensation of boson pairs in momentum space
(identical to the Cooper condensate of fermion pairs \cite{23}).

The degeneracy of equation (\ref{26}) with respect to the phase of
$\tilde\Psi_{12}(p)$ at $n_0\to 0$ allows for
the condition of stability of the phonon spectrum
$c_1^2=\tilde\Psi_{12}(0)/\tilde m^*>0$ to be met
by means of choosing the appropriate sign (phase) of the pair
order parameter $\tilde\Psi_{12}(0)>0$.
However, the model of the SF state with PCC and no BEC \cite{21}--\cite{22}
implies a few paradoxes, such as a finite energy gap in the single-particle
spectrum at $p=0$, exponential asymptotic behavior of
the pair correlation function
$\left\langle \hat \psi (\mathbf{r}) \hat \psi (\mathbf{r'}
)\right\rangle $ at $|\mathbf{r}-\mathbf{r'}|\to \infty$,
half-integer quantum of circulation of the SF velocity
$\kappa=\hbar/2m$ etc.

Indeed, at $\mathbf{p}\to 0$ and
$n_0 \ne 0$,
Eq.~(\ref{26}), due to  the isotropic momentum
dependence of the spectrum $E(p)$ and the functions
$\tilde V(p)\equiv \tilde V(\mathbf{p}, E(p))$,
$\Lambda (p)\equiv \Lambda (\mathbf{p}, E(p))$ and
$\tilde \Psi_{12}(p)\equiv
\tilde \Psi_{12}(\mathbf{p}, E(p))$  reduces to
the form
\begin{equation}
\tilde \Psi_{12}(0)=- \frac{1}{(2\pi )^2}\int\limits_{0}^{\infty}
\frac{ k^2dk}{E(k)}
\left[ n_0\Lambda^2(k)\tilde V^2(k)+\Lambda(k)
\tilde V(k) \tilde \Psi_{12}(k)\right]\;.
\label{30}
\end{equation}
The first integral addend on the right-hand side
of Eq.~(\ref{30}) being always negative, the value of
$\tilde \Psi_{12}(0)$ can be negative as well.
The condition $\tilde \Psi_{12}(0)<0$ means that the phase of
the PCC is opposite to phase of the BEC, because $n_0>0$.
Moreover, in this case, in spite of the condition $\Lambda(0) \tilde V(0)>0$
(which ensures that the system is globally stable against a
spontaneous collapse), at sufficiently small densities of the BEC,
in accordance with Eq.~(\ref{20}), the values
\begin{equation}
\tilde \Sigma_{12}(0,0)=n_0\Lambda(0)\tilde V(0)-\vert
\tilde \Psi_{12}(0)\vert
\label{31}
\end{equation}
become negative if
$\vert \Psi_{12}(0)\vert >n_0\Lambda (0) V(0)$,
which corresponds to an
instability in the phonon spectrum ($ c_1^2<0$).
However, if the pair interaction between bosons in a broad enough
region of the momentum space has the
character of attraction, i.e.,
$\Lambda(k) \tilde V(k)<0$ at $k\ne0$,
and if the magnitude of that attraction is
large enough (see below),
the second (positive) addend on the right-hand side
of Eq.~(\ref{30}) can outweigh the first (negative) one
if the BEC density is small enough $(n_0\ll n)$. Then
$\tilde \Psi_{12}(0)$ will be positive, and the phase
of the PCC will coincide with phase of the BEC,
so that $\tilde \Sigma_{12}>0$
and $c_1^2>0$.

Since at $T=0$ the density of the SF component
$\rho_s$, on the one hand, coincides with the total density $\rho=mn$
of the Bose liquid and, on the other hand, $\rho_s$ is proportional
to $\tilde\Sigma_{12}(0,0)$ which plays the role of the SF order parameter,
with account for (\ref{20}) and (\ref{30}),
one
gets the following relations:
\begin{equation}
\rho_s\equiv \rho_0+\tilde \rho_s=
\beta m \frac{\tilde\Sigma_{12}(0,0)}{\Lambda(0)\tilde V(0)}
=\beta m
\left[n_0(1-\gamma)+\Psi\right]
\label{32}
\end{equation}
where
\begin{equation}
\gamma=\frac{1}{(2\pi)^2\Lambda(0)\tilde V(0)}\int\limits_{0}^{\infty}
\frac{k^2dk}{E(k)}\left[\Lambda(k)\tilde V(k)\right]^2,
\label{33}
\end{equation}
\begin{equation}
\Psi=-\frac{1}{(2\pi)^2\Lambda(0)\tilde V(0)}\int\limits_{0}^{\infty}
\frac{k^2dk}{E(k)}\Lambda (k)\tilde V(k) \tilde\Psi_{12}(k)\;,
\label{34}
\end{equation}
and $\beta$ is a certain dimensionless constant.
Since the density of the single-particle BEC is equal to $\rho_0=mn_0$,
we obtain $\beta=\left( 1-\gamma\right)^{-1}.$
This means that the density of the ``Cooperlike'' PCC is
\begin{equation}
\tilde\rho_s=mn_1=\frac{m\Psi}{(1-\gamma)}\;,
\label{35}
\end{equation}
where the
concentration $n_1=n-n_0$ is then determined from relation
(\ref{29}), and for liquid $^4$He at $T\to 0$, in accordance with
the experimental data \cite{15}--\cite{19}, it should be approximately
$90\%$ of the full concentration of $^4$He atoms in
liquid helium $n=2.17\cdot 10^{22}$~cm$^{-3}$.
Thus, the SF component  of Bose liquid in this model at $T=0$
is an effective
coherent condensate \cite{12} which is  a superposition of the
weak one-particle BEC and intensive PCC.

\section{Choice of the pair interaction potential in the Bose liquid}
To describe interaction of helium atoms in real space,
various semi-empirical potentials
are conventionally used, which describe
strong repulsion at small distances and weak van der Waals
attraction at large distances.
However, most of those potentials are characterized by a strong
divergence at $r\to 0$, like, for instance, the Lennard-Jones potential
\begin{equation}
U_\mathrm{LJ}(r)=
\epsilon\left[\left( \frac{\sigma}{r}\right)^{12}-
\left(\frac{\sigma}{r}\right)^{6}\right],\qquad r>r_c\;.
\label{37}
\end{equation}
Such potentials are not suitable for the
description of pair interaction in momentum space, since the
respective Fourier components
\begin{equation}
V(p)=\int d^3r\,U(r)\exp{(i\mathbf{p} \mathbf{r})}=
\frac{4\pi}{p}\int\limits_{0}^{\infty} r U(r)\sin{(p r)}\,dr
\label{38}
\end{equation}
are infinite, diverging at the lower limit.
Lately, in the calculations of interatomic interaction
and possible bound states, i.e., He$_2$ molecules,
one uses more up-to-date potentials, like the Aziz potential \cite{3}--\cite{5}:
\begin{eqnarray}
&& U_\mathrm{A}(r) =
A\exp(-\alpha r - \beta r^2) - F(r,r_0)
(c_6r^{-6} + c_8r^{-8} + c_{10}r^{-10})\;,
\nonumber \\
&& F(r,r_0) = \left\{
\begin{array}{ll}
\exp\left[-(r_0/r-1)^2\right]\sum_{k=0}^2\;,& r < r_0\,\\
1\;,& r \ge r_0
\end{array}
\right.
\label{Aziz}
\end{eqnarray}
where $A=1.8443101\times10^5$~K, $\alpha=10.43329537$~\AA$^{-1}$,
$\beta=2.27965105$~\AA$^{-2}$, $c_6 = 1.36745214\mbox{ K}\times\mbox{\AA}^6$,
$c_8 = 0.42123807\mbox{ K}\times\mbox{\AA}^8$,
$c_{10} = 0.17473318\mbox{ K}\times\mbox{\AA}^{10}$.
Such potentials  remain finite at $r=0$ due to the nonanalytic
exponential $r$ dependence, which suppresses any
power divergence at $r\to0$ (see Fig.~3).
However, employing its Fourier component in solving
the nonlinear integral equations (\ref{25})--(\ref{26})
is technically difficult while not conclusive
by itself: In an intrinsically many-body problem like
the one at hand, collective effects are certain to
play an essential part and to render the
subtleties in the shape of the two-body potential
largely irrelevant. In order to retain the crucial
features of the system yet not be overwhelmed by
technical complications, we will utilize a model potential
of ``soft spheres'' describing
(unlike the ``hard spheres'' model \cite{28}) finite repulsion in
a certain bound region, which accounts for effects of
mutual quantum diffraction of bosons in the Bose liquid.

In this context, consider a model potential in the form of a Fermi-type
function in real space (Fig.~4a)
\begin{equation}
U_\mathrm{F}(r)=U_0\left\{\exp \left(
\frac{r^2-a^2}{b^2}\right)+1\right\}^{-1}\;,
\label{39}
\end{equation}
which at $b=0$ degenerates into a ``step'' of
finite height $U_0$ at $r<a$.
In this latter case the Fourier component is expressed in terms of the
first order spherical Bessel function (see Fig.~5):
\begin{equation}
V(p)=V_0\frac{j_1(pa)}{pa};\qquad
j_1(x)=\frac{\sin(x)-x \cos(x)}{x^2}\;.
\label{40}
\end{equation}
where $V_0\equiv 3V(0)=4\pi U_0a^3$.
The same oscillating Fourier component is characteristic of
a smooth potential $V(r)$ in the
form of a Lindhardt-type  function \cite{23},  having an infinite
negative derivative at the inflection point $r=a$ (see Fig.~4b):
\begin{equation}
U_L(r)=
\frac{U_0}{2}\left[1+\frac{\left( 1-r^2/a^2\right)}
{2r/a}\ln{\left|\frac{a+r}{a-r}\right|}
\right].
\label{41}
\end{equation}
Formally, this problem is an inverse to the one of
periodic oscillations of spin density in real space
of interacting spins in metal (so-called
Ruderman-Kittel-Kasui-Yoshida oscillations \cite{Whiyte}).

Note that the amplitude of
oscillations of the Fourier component for the Fermi type potentials
(\ref{39}) at $b\ne 0$ is damping exponentially with the increase
of the parameter $b$, due to the decreasing absolute value
of the negative derivative at the inflection point
(see Fig.~5, inset).

Such oscillations of the Fourier component of
the pair potential $U(r)$ in momentum space arise even in the
absence of attraction in real space and are a consequence of
quantum diffractional effects of mutual scattering
of the particles.
This means that the existence of negative values $V(p)<0$,
i.e., of effective attraction in some regions of momentum
transfer $p$, is not directly associated with
van der Waals forces, which are explicitly taken into account
in the sign-changing (with respect to $r$) Lennard-Jones or
Aziz potentials (see Fig.~3).

If one substitutes the oscillating potential (\ref{40})
into the Bogolyubov spectrum of a dilute quasi-ideal Bose gas \cite{24}
\begin{equation}
E_B(p)=\left\{\frac{p^2}{2m}\left[\frac{p^2}{2m}+
2nV(p)\right]\right\}^{1/2},
\label{42}
\end{equation}
then, by choosing two parameters, $V_0$ and $a$,
independently, one can achieve a rather satisfactory coincidence
of the spectrum $E_B(p)$ with the elementary excitation
spectrum $E_\mathrm{exp}(p)$ in liquid $^4$He derived from
neutron scattering experiments \cite{30}--\cite{32}
(Fig.~6, solid and dotted curves).
However, the spectrum (\ref{42}) with the potential (\ref{40})
turns out to be unstable at large values of $V_0$,
because $E_B^2(p)<0$ in some range of $p$
(Fig.~6, dashed curve).
Moreover, the Bogolyubov model of the quasi-ideal Bose gas
with an intensive BEC at $T\to 0$ ($n_0\to n$) is not applicable
to the description of the Bose liquid with a strongly suppresed BEC
($n_0\ll n$).

On the other hand, multiparticle collective effects in the Bose
liquid, according to Eq.~(\ref{22}), lead to an essential
renormalization of the pair interaction, which determines the
normal and anomalous self-energy parts, Eqs.~(\ref{19}) and (\ref{20}).
An important feature of the renormalized interaction (\ref{22})
is that in the regions of phase volume $(\mathbf{p}, \omega)$
where the real part of $\Pi (\mathbf{p}, \omega)$ is negative,
the repulsion (when $V(p)>0$) gets suppressed while the attraction
(when $V(p)<0$) gets effectively enhanced.
This fact was first noted in Ref.~\cite{33} and used in Ref.~\cite{34}
where the integral equations (\ref{25}) and (\ref{26}) were solved
with the seed potential of the hard spheres model (see (\ref{37})).
Note that in Ref.~\cite{34}, a possibility for bound
pairs of helium atoms to form not only in momentum space,
but also in real space
was discussed, which allowed one to interpret the anomalously large
effective mass $m_3^*$ of the dope $^3$He atoms in the
SF Bose liquid $^4$He \cite{34-1}--\cite{34-2} as the mass
of a bound $^3$He-$^4$He pair, equal to $M=m_3+m_4=7 m_3/3$.

In this paper, we take into account the explicit momentum
dependence of the real part of polarization operator
$\Pi(\mathbf{p},E(p))$, which can be represented as
(see Appendix):
\begin{equation}
\begin{array}{c}
\displaystyle
\Re \,\Pi(\mathbf{p}, E(p))=\frac{1}{2}
\int\frac{d^3\mathbf{k}}{(2\pi)^3}
\frac{\Gamma(\mathbf{p},\mathbf{k})}
{E(k)-E(|\mathbf{k}-\mathbf{p}|)-E(p)} \\[12pt]
\displaystyle
{}\times\left\{\frac{F_{-}(\mathbf{k}, \mathbf{p})}
{E(k)[E(k)+E(|\mathbf{k}-\mathbf{p}|)-E(p)]} -
\frac{F_{+}(\mathbf{k}, \mathbf{p})}
{E(k)[E(k)+
E(|\mathbf{k}-\mathbf{p}|)+E(p)]}\right\}
\label{44}
\end{array}
\end{equation}
where
\begin{equation}
\begin{array}{c}
\displaystyle
F_{-}(\mathbf{k}, \mathbf{p})=\left[
E(k)+\frac{\mathbf{k}^2}{2m}-\mu+
\tilde \Sigma_{11}(\mathbf{k}, E(k))\right]
\\[12pt]
\displaystyle{}\times\left[
E(k)-E(p)+\frac{(\mathbf{k}-\mathbf{p})^2}{2m}-\mu+
\tilde \Sigma_{11}(\mathbf{k}-\mathbf{p},
E(k)-E(p))\right]\\[12pt]
{}+\tilde \Sigma_{12}(\mathbf{k}, E(k))
\tilde \Sigma_{12}(\mathbf{k}-\mathbf{p}, E(k)-E(p))\;,
\label{45}
\end{array}
\end{equation}

\begin{equation}
\begin{array}{c}
\displaystyle
F_{+}(\mathbf{k}, \mathbf{p})=\left[
E(|\mathbf{k}-\mathbf{p}|)+\frac{(\mathbf{k}-\mathbf{p})^2}{2m}-\mu+
\tilde \Sigma_{11}(\mathbf{k}-\mathbf{p}, E(|\mathbf{k}-\mathbf{p}|))\right]
\\[12pt]
\displaystyle
{}\times\left[
E(|\mathbf{k}-\mathbf{p}|)+E(p)+\frac{\mathbf{k}^2}{2m}-\mu+
\tilde \Sigma_{11}(\mathbf{k},
E(|\mathbf{k}-\mathbf{p}|)+E(p))\right]
\\[12pt]
{}+\tilde \Sigma_{12}(\mathbf{k}-\mathbf{p},
E(|\mathbf{k}-\mathbf{p}|))
\tilde \Sigma_{12}(\mathbf{k}, E(|\mathbf{k}-\mathbf{p}|)+E(p))\;.
\label{46}
\end{array}
\end{equation}

As follows from Eq.~(\ref{44}), if  the quasiparticle spectrum
$E(p)$ is
stable with respect to decays into a pair of quasiparticles
\cite{7},\cite{37}, i.e., if for all $\mathbf{p}$ and $\mathbf{k}$ the
following
conditions  are fulfilled:
\begin{equation}
E(p)<E(
k)+E(|\mathbf{k}-\mathbf{p}|)\;; \qquad
E(k)<E(p)+E(|\mathbf{k}-\mathbf{p}|),
\label{47}
\end{equation}
the common denominator in
front of the curly braces is always negative,
\begin{equation}
E(k)-E(|\mathbf{k}-\mathbf{p}|)-E(p)<0\;,
\label{48}
\end{equation}
whereas the denominator of the first term in the curly braces is
always positive,
\begin{equation}
E(k)+E(|\mathbf{k}-\mathbf{p}|)-E(p)>0
\label{49}
\end{equation}
and smaller than the positive denominator of the second term
\begin{equation}
E(k)+E(|\mathbf{k}-\mathbf{p}|)+E(p)>0\;.
\label{50}
\end{equation}
This means that the integrand of Eq.~(\ref{44})
is negative if the functions $F_{\pm}(\mathbf{k},\mathbf{p})$
are positive.
According to the numerical calculations \cite{34}, \cite{PMV}
in the framework of the ``hard'' and ``soft'' sphere models,
$F_{\pm}(\mathbf{k},\mathbf{p})>0$ for all
$\mathbf{k}$ and $\mathbf{p}$, so that the real part of $\Pi(\mathbf{p},
E(p))$ is negative, because $\Gamma>0$ (see below).
One should note that the actual experimental spectrum of elementary
excitations in liquid $^4$He is decaying at small enough
momenta.
However, this will not change the negative sign of $\Pi(\mathbf{p})$
due to the integral character of expression (\ref{44})
(cf.~\cite{PMV}).
Note also that here we do not take into account the imaginary part
of $\Pi(\mathbf{p},\omega)$, which determines
the damping of quasiparticles and the dynamical structure factor
(see Appendix).

\section{The iterative scheme of calculation of the quasiparticle
spectrum}

In order to calculate the quasiparticle spectrum within the model
of a Bose liquid with a suppressed BEC and intensive
PCC being considered,
at first, using Eqs.~(\ref{25}) and (\ref{26}), a numerical
calculation in the first approximation of the functions
$\Phi_1(p)\equiv\tilde \Psi_{11}(\mathbf{p},E_0(\mathbf{p}))$ and
$\Psi_1(p)\equiv \tilde\Psi_{12}(\mathbf{p},E_0(\mathbf{p}))$, was
conducted. For the zeroth approximation, the Bogolyubov spectrum
$E_0(p)=E_B(p)$ and the ``screened'' potential (\ref{22}) with
account for Eq.~(\ref{40}) were taken:
\begin{equation}
\tilde V_0(p)=\frac{V_0j_1(pa)}{pa-V_0\Pi_0 j_1(pa)}
\label{51}
\end{equation}
at some constant negative value of $\Pi_0$.
Then, using the functions $\Phi_1(p)$ and $\Psi_1(p)$
obtained, the first approximation for the polarization
operator $\Pi_1(p)$ was calculated, using Eqs.~(\ref{44})--(\ref{46})
at $\Gamma=1$. Here, too, the Bogolyubov spectrum (\ref{42}),
which is the best fit to the empiric spectrum $E_\mathrm{exp}(p)$
for liquid $^4$He (see Fig.~6) was chosen as the zeroth approximation
for $E(p)$.
The limiting value $\Pi_1(0)$ was compared with the exact
thermodynamic value of the polarization operator of
the $^4$He Bose liquid at $p=0$ and $\omega=0$ [cf.~(\ref{1})],
which determines the compressibility of the Bose system:
$\Pi(0,0)=-n/mc_1^2$.

The absolute value $\vert\Pi(0,0) \vert$ turned out to be
almost 1.5 times greater than the calculated value $\vert\Pi_1(0)\vert$.
This provides an estimate of the vertex $\Gamma$ at $p=0$
in the first approximation as $\Gamma_1\equiv\Lambda_1\simeq 1.5$.
The second approximation $\Phi_2(p)$ and $\Psi_2(p)$ was obtained
from Eqs.~(\ref{25}), (\ref{26}) with the constant value
$\Gamma_1\equiv\Lambda_1$ and the first approximation for the
renormalized potential, Eq.~(\ref{44}):
\begin{equation}
\tilde V_1(p)=\frac{V_0j_1(pa)}{pa-V_0\Pi_1(p)j_1(pa)}\;.
\label{52}
\end{equation}
Such an iterative procedure was repeated four to six times
and used to improve precision in the calculation of the
polarization operator. At each stage, equations (\ref{27})
and (\ref{28}) were used to reproduce the quasiparticle
spectrum $E(p)$, and the rate of convergence of the iterations
was watched, as well as the
 degree of proximity of $E(p)$
to the empirical spectrum $E_\mathrm{exp}(p)$.

The only fitting parameter in these calculations was the
amplitude $V_0$ of the seed potential (\ref{40})
at the value of $a=2.44$~\AA, which is equal to twice the
quantum radius of the $^4$He atom. The BEC density,
in accordance with the experimental data \cite{15}--\cite{19},
was fixed at $n_0=9\% n=1.95\cdot 10^{21}\,\mathrm{cm}^{-3}$.
The computation has resulted in a quite satisfactory agreement
of the theoretical spectrum $E(p)$ with $E_\mathrm{exp}(p)$.
Figure 8 depicts the momentum dependence of $\Pi(p)$ as obtained
with five iterations, while Fig.~9 shows the self-consistent $p$
dependences of $\Phi(p)$, $\Psi(p)$, and $A(p)$ obtained
from Eqs.~(\ref{25}), (\ref{26}), and (\ref{28}). One notices
that the functions $\Phi(p)$ and $\Pi(p)$ are negative at all $p$,
whereas $\Psi(p)$ and $A(p)$ are positive. The locations of
the deep minima of $\Phi(p)$ and $A(p)$ practically coincide with
the location of the minimum of the potential
(\ref{22}) (see Fig.~7).

Finally, in Fig.~10, the solid curve is the theoretical
quasiparticle spectrum $E(p)$ obtained from Eq.~(\ref{27}),
and the dots are the experimental spectrum obtained from
data on inelastic neutron scattering in $^4$He \cite{30}--\cite{32}.
In the calculation of $E(p)$,  the  fitting parameter
$V_0$ was chosen in such a way that the phase velocity
of quasiparticles $E(p)/p$ at $p\to 0$ coincide with the
speed of the first hydrodynamic sound $c_1\simeq 236$~m/s
in liquid $^4$He. This value of $V_0$ corresponds to
the repulsion potential of the ``soft'' spheres model
$U_0=V_0/(4\pi a^3)=1552$~K at $a=2.44$~\AA.
We see that there is a satisfactory agreement of $E(p)$ with
$E_\mathrm{exp}(p)$ in the region $p\le 2.5 \mbox{ \AA}^{-1}$.
For $p>2.5 \mbox{ \AA}^{-1}$, the theoretical spectrum $E(p)$
lies somewhat higher than $E_\mathrm{exp}(p)$, which,
apparently, has to do with the fact that the vertex function
$\Gamma(\mathbf{k}, \mathbf{p})$, a decreasing function of $p$,
was replaced with a constant value
$\tilde \Gamma\simeq 1.5$ for all $p$.

Of course, the value of $E(p)$ at large momenta should not exceed
the doubled value of the roton gap $\Delta_r=8.61$~K lest
the spectrum becomes decaying \cite{37}.
To check this, we have approximated the vertex $\Lambda(p)=\Gamma(0,p)$
with a slowly decreasing function, falling
down from $\tilde \Gamma= 1.5$ to $\tilde \Gamma= 1.1$
on the interval $2.1\mbox{ \AA}^{-1}<p<3.8\mbox{ \AA}^{-1}$.
The resulting theoretical spectrum is shown in Fig.~11,
together with the experimental data
\cite{30}--\cite{32} (light circles) and the latest
results \cite{Pearce}  (asterisks) of measurements of the
spectrum at $2\mbox{ \AA}^{-1}<p<3.6 \mbox{ \AA}^{-1}$,
beyond the roton minimum.
Evidently, such an approximation for the vertex part
yields a much better agreement of the theoretical
and experimental spectra at large values of momentum.
Note also that the self-consistency of this model
is corroborated by the fact
that the theoretical value of total particle concentration
calculated from Eq.~(\ref{29}),
$n_\mathrm{th}=2.12\cdot 10^{22}\mbox{ cm}^{-3}$,
is quite close to the experimental value
for liquid $^4$He, $n=2.17\cdot 10^{22}\mbox{ cm}^{-3}$
(at $n_0=9\%\,n$).
On the other hand, the concentration $n_1$ of
supracondensate particles,
calculated from Eqs.~(\ref{33})--(\ref{35})
at the values of
the parameters indicated, is about $0.93\,n$, which is also in
good accordance with experiment, taking into account that
the BEC density is determined up to $\pm 0.01\,n$.
Also, when formulating the approximate theoretical model,
quadratic terms in the small parameter $n_0/n\ll 1$
in Eqs.~(\ref{19})--(\ref{20}) were omitted, which also introduces an
error of the order of $1\%$ \cite{15}--\cite{20}.

\section{Conclusions}

Thus, the model of the SF state of a Bose liquid with
a single-particle BEC suppressed because of interaction
and an intensive PCC, based upon a renormalized field perturbation
theory  with combined variables \cite{11}--\cite{13},
allows one to obtain a self-consistent ``trimmed'' system of
nonlinear integral equations for the self-energy parts
$\tilde\Sigma_{ij}(p,\epsilon)$, by means of truncating the
infinite series in the small density of the BEC ($n_0/n\ll 1$).
By the same token, one can work out a self-consistent microscopic
theory of a superfluid Bose liquid and perform an ab initio
calculation of the spectrum of elementary excitations $E(p)$,
starting from realistic models of pair interaction potential
$U(r)$ possessing finite Fourier components.
It is shown that for a  repulsive potential
in the framework of  the ``soft spheres'' model,
the Fourier component $V(p)$ is an oscillating sign-changing
function of momentum transfer $p$. This means  that in certain
regions of momentum space at $p\ne 0$ there is an effective
attraction between bosons, $V(p)<0$, which has nothing to do
with van der Waals forces and has a quantum mechanical diffraction
nature. That attraction gets substantially enhanced due to
multiparticle collective effects of renormalization (``screening'') of
the initial interaction, which are described by the boson polarization
operator $\Pi(\mathbf{p},\omega )$.
The enhancement of the attraction happens because on the
``mass shell'' $\omega=E(p)$, the real part of
$\Pi(\mathbf{p},E(p))$ is
negative in the whole region of momentum where  the quasiparticle
spectrum $E(p)$ is stable with respect to decay \cite{37}.
It is necessary to emphasize that this negative sign
of $\Re\,\Pi (\mathbf{p}, E(p))$
is only characteristic of Bose systems,  in which
the single-particle and collective spectra coincide with each other
and are measured from the
common zero of energy, unlike the Fermi systems,
in which the single-particle
excitation spectrum begins at the Fermi energy,
due to the Pauli principle.
Therefore, in the $^3$He Fermi liquid there can
be no corresponding effective enhancement of the
negative values of the same ``input'' interaction potential $V(p)$,
so that the formation of Cooper pairs
is only possible for nonzero orbital momenta,
due to the true weak van der Waals attraction between fermions
\cite{Pitaevskiy}--\cite{Anderson}.
Apparently, it is this fact that
has to do with the critical temperatures
of the SF transition in $^4$He and $^3$He
differing by three orders of magnitude.

The rather strong pair
attraction of bosons in momentum space
for $\Re\,\Pi(p,\omega)<0$ forms an
intensive PCC, which, together with a weak BEC,
constitutes a single coherent condensate, making up
the microscopic foundation of the SF component of the Bose
liquid $\rho_s\sim \tilde\Sigma_{12}(0,0)$.
On the other hand, the
oscillating nature of the renormalized Fourier
component of the potential
$\tilde V(p)$ (see Fig.~5 and Fig.~7) leads to
a nonmonotonic behavior of
momentum dependences of the mass operators
$\tilde \Sigma_{11}(p, E(p))$ and $\tilde \Sigma_{12}(p, E(p))$,
and, as a consequence, to the emergence of a roton minimum in
the quasiparticle spectrum $E(p)$, which is directly connected
with the deepest first negative minimum of $\tilde V(p)$.

It is necessary to emphasize that
the amplitude of the ``soft spheres'' model repulsion potential
obtained, $U_0=1552$~K, which corresponds to very good agreement
between the theoretical quasiparticle spectrum $E(p)$ and
the experimental spectrum of elementary excitations in
$^4$~He is smaller than the
value of the Aziz type potential at $r\to 0$ (see Fig.~3, inset).
It is a result of strong quantum diffraction effects
in the Bose liquid, because the average distance between
particles is of the order of or less than the de Broglie wavelength
of the bosons.

\section{Acknowledgments}
We are grateful to P.I.~Fomin, I.V.~Simenog,
E.Ya.~Rudavsky, I.N.~Adamenko, L.V.~Karnatsevich, M.A.~Strzhemechny,
and S.I.~Shevchenko for useful discussions of theoretical and
experimental aspects of the model of SF liquid under consideration here.

\section{Appendix}
The polarization operator (\ref{19}) can be
calculated without account
for the vertex part $\Gamma$, making use of
expressions (\ref{3})--(\ref{6}) in the form
$$
\Pi(\mathbf{p},\omega)=\int \frac{d^3\mathbf{k}}{(2\pi)^3}\left[L_{11}
(\mathbf{p}, \mathbf{k}, \omega)+L_{12}(\mathbf{p}, \mathbf{k}, \omega)
\right]\;,
\eqno(\mathrm{A}.1)$$
where
$$
L_{ij}(\mathbf{p}, \mathbf{k},
 \omega)=
i\oint \frac{dz}{2\pi} \tilde{G}_{ij}(\mathbf{k}, z)
\tilde{G}_{ij}(\mathbf{k}-\mathbf{p}, z-\omega)\;.
\eqno(\mathrm{A}.2)$$
Assume that the Green functions $\tilde{G}_{ij}$ have only one pole
within the integration contour and are equal to
$$
\displaystyle
\tilde{G}_{11} (\mathbf{k}, \epsilon) =
\frac{\epsilon+\frac{k^2}{2m}-\mu+\tilde{\Sigma}_{11}(-\mathbf{k},
 -\epsilon)}
{\epsilon^2-E^2(k)+i\delta}\;;
\eqno(\mathrm{A}.3)$$
$$
\displaystyle
\tilde{G}_{12} (\mathbf{k}, \epsilon) =
\frac{\tilde{\Sigma}_{12}(\mathbf{k}, \epsilon)}
{\epsilon^2-E^2(k)+i\delta}\; \qquad (\delta\to 0)\;.
\eqno(\mathrm{A}.4)$$
Calculating the integrals (A.2) with account for the poles at the points
$\epsilon=E(k)$ and $\epsilon=E(|\mathbf{k}-\mathbf{p}|)+\omega$ yields
$$\begin{array}{c}
\displaystyle
L_{11}(\mathbf{p}, \mathbf{k}, \omega)=
\frac{1}{2\left[
E(k)-E(|\mathbf{k}-\mathbf{p}|)-\omega\right]}
\left\{
\left[E(k)+\frac{\mathbf{k}^2}{2m}-\mu+\tilde{\Sigma}_{11}
(\mathbf{k}, E(k))\right]\right.
\\[20pt]
\displaystyle
\left.{}\times\frac{
\left[
E(k)-\omega+\frac{(\mathbf{k}-
\mathbf{p})^2}{2m}-\mu+\tilde{\Sigma}_{11}
(\mathbf{k}-\mathbf{p}, E(k)-\omega)\right]}
{E(k)\left[
E(k)+E(|\mathbf{k}-\mathbf{p}|)-\omega\right]}
\right.\\[20pt]
\displaystyle
\left.{}-\left[
E(|\mathbf{k}-\mathbf{p}|)+\frac{(\mathbf{k}-
\mathbf{p})^2}{2m}-\mu+\tilde{\Sigma}_{11}
(\mathbf{k}-\mathbf{p}, E(|\mathbf{k}-\mathbf{p}|))
\right]\right.
\\[20pt]
\displaystyle
\left.{}\times
 \frac{\left[
E(|\mathbf{k}-\mathbf{p}|)+\omega+\frac{\mathbf{k}^2}{2m}
-\mu+\tilde{\Sigma}_{11}
(\mathbf{k}, E(|\mathbf{k}-\mathbf{p}|)+
\omega)\right]}
{E(|\mathbf{k}-\mathbf{p}|)\left[
E(k)+E(|\mathbf{k}-\mathbf{p}|)+\omega\right]}\right\}\;,
\end{array}
\eqno(\mathrm{A}.5) $$
$$
\begin{array}{c}
\displaystyle
L_{12}(\mathbf{p}, \mathbf{k}, \omega)=
\frac{1}{2\left[
E(k)-E(|\mathbf{k}-\mathbf{p}|)-\omega\right]}
\left\{
\frac{\tilde \Sigma_{12}(\mathbf{k},E(k))
\tilde \Sigma_{12}(\mathbf{k}-\mathbf{p},E(k)-\omega)}
{E(k)\left[
E(k)+E(|\mathbf{k}-\mathbf{p}|)-\omega\right]}
\right.\\[20pt]
\displaystyle
\left.{}-
\frac{\tilde \Sigma_{12}(\mathbf{k}, E(|\mathbf{k}-\mathbf{p}|)+\omega)
\tilde \Sigma_{12}(\mathbf{k}-\mathbf{p}, E(|\mathbf{k}-\mathbf{p}|))}
{E(|\mathbf{k}-\mathbf{p}|)\left[E(k)+E(|\mathbf{k}-\mathbf{p}|)
+\omega\right]}
\right\}
\end{array}
\eqno(\mathrm{A}.6)$$

In the statistical limit $(\omega\to 0,\mathbf{p}\to 0)$, expression (A.5)
reduces to
$$
\begin{array}{c}
\displaystyle
L_{11}(0,\mathbf{k},0)=-\frac{1}{4}\left\{\frac{1}{E^2(k)}\left[
\epsilon(k) +\frac{\mathbf{k}^2}{2m}-\mu+
\tilde \Sigma_{11}(\mathbf{k}, \epsilon(k))\right]^2\right.
\\[12pt]
\displaystyle
\left.
{}+\left[
\frac{2}{\epsilon(k)}\left(1+\frac{\partial \tilde \Sigma_{11}(\mathbf{k})}
{\partial \epsilon}\right)
-\frac{k}{m\epsilon(k)}
\frac{1}{\frac{\partial\epsilon(k) }{\partial k}}\right]
\left[
\epsilon(k) +\frac{\mathbf{k}^2}{2m}-\mu+
\tilde \Sigma_{11}(\mathbf{k}, \epsilon(k))\right]\right\}
\end{array}
\eqno(\mathrm{A}.7)$$

It follows that in a large region of momentum space,
$I_{11}(0,\mathbf{k},0)<0$.
 The same result is obtained for the
function (A.6) at $p=0$ and $\omega=0$, i.e., $L_{12}(0,\mathbf{k},0)<0$,
so that the static bosonic polarization operator $\Pi(0,0)$ is negative,
which corresponds to a suppression of the ``screened'' repulsion at
$\mathbf{p}\to 0$. From Eqs.~(A.5) and (A.6) it can also be seen that on
the ``mass shell'' $\omega=E(p)$, the integrals $L_{11}$ and $L_{12}$
remain negative in a wide region of momentum space because of the
negative sign of the common denominator
$E(k)-E(|\mathbf{k}-\mathbf{p}|)-E(p)<0$
and the positive sign of the denominator
$E(k)+E(|\mathbf{k}-\mathbf{p}|)-E(p)>0$
due to the fact that the quasiparticle spectrum $E(p)$ is decayless
[conditions (\ref{47})].

Thus, the real part of the polarization operator (A.1) at  $\omega=E(p)$
is negative on the whole range of $p$.
In order to determine the imaginary part of $\Pi(\mathbf{p}, \omega)$,
which describes Landau quantum damping of bosons, one has to calculate
the main value of the following integral:
$$
\begin{array}{c}
\displaystyle
L(\mathbf{p}, \mathbf{k}, \omega)=
\frac{i}{2\pi}
\;\mbox{V.p.}\;\int\limits_{-\infty}^{\infty}\frac{d\epsilon}{\left[
\epsilon^2-E^2(k)\right]
\left[
(\epsilon-\omega)^2-E^2(\mathbf{k}-\mathbf{p})\right]}\\[12pt]
\left\{\left[\epsilon+\sigma(\mathbf{k},\epsilon)\right]
\left[\epsilon-\omega+\sigma(\mathbf{k}-\mathbf{p},\epsilon-\omega)\right]
+\tilde\Sigma_{12}(\mathbf{k},\epsilon)\tilde\Sigma_{12}
(\mathbf{k}-\mathbf{p},\epsilon-\omega)
\right\},
\end{array}
\eqno(\mathrm{A}.8)$$
where
$$
\sigma(\mathbf{k},\epsilon)=\frac{\mathbf{k}^2}{2m}-\mu+
\tilde\Sigma_{11}(\mathbf{k},\epsilon)
$$

Factoring the denominators in the integrand of(A.8)
into simple fractions, one obtains
$$
\begin{array}{c}
\displaystyle
L(\mathbf{p}, \mathbf{k}, \omega)=
\frac{i}{2\pi E(|\mathbf{k}-\mathbf{p}|)}
\;\mbox{V.p.}\;\int\limits_{-\infty}^{\infty}d\epsilon\left[
M_+(\mathbf{p}, \mathbf{k}, \omega,\epsilon)-
M_-(\mathbf{p}, \mathbf{k}, \omega,\epsilon)
\right]
\end{array}
\eqno(\mathrm{A}.9)
$$
where
$$
\begin{array}{c}
\displaystyle
M_{\pm}(\mathbf{p}, \mathbf{k}, \omega,\epsilon)=
\frac{\epsilon C_{\pm}(\mathbf{p}, \mathbf{k}, \omega,\epsilon)+
D_{\pm}(\mathbf{p}, \mathbf{k}, \omega,\epsilon)}
{ \epsilon^2-E^2(k)}+
\frac{C_{\pm}(\mathbf{p}, \mathbf{k}, \omega,\epsilon)}
{\epsilon-\omega\mp E(|\mathbf{k}-\mathbf{p}|)}\;;
\end{array}
\eqno(\mathrm{A}.10)
$$

$$
\begin{array}{c}
\displaystyle
C_{\pm}(\mathbf{p}, \mathbf{k}, \omega,\epsilon)=
\frac{1}
{E^2(k)-\left[\omega\pm E(\mathbf{k}- \mathbf{k})\right]^2}
\left\{
\left[E(k)+\sigma(\mathbf{k},\epsilon)\right]
\sigma(\mathbf{k}-\mathbf{p},\epsilon-\omega
)
\right.
\\[12pt]
\displaystyle
\left.
{}+
\left[E(|\mathbf{k}-\mathbf{p}|)+\omega\right]
\left[\sigma(\mathbf{k}-\mathbf{p},
\epsilon-\omega) -\omega
\right]+
\tilde\Sigma_{12}(\mathbf{k},\epsilon)\tilde\Sigma_{12}
(\mathbf{k}-\mathbf{p},\epsilon-\omega)
\right\}\;;
\end{array}
\eqno(\mathrm{A}.11)
$$

$$
\begin{array}{c}
\displaystyle
D_{\pm}(\mathbf{p}, \mathbf{k}, \omega,\epsilon)=
\sigma(\mathbf{k},\epsilon)+
\sigma(\mathbf{k}-\mathbf{p},\epsilon-\omega)-\omega
+\left[\omega\pm E(|\mathbf{k}-\mathbf{p}|)\right]
C_{\pm}(\mathbf{p},\mathbf{k},\epsilon,\omega)\;.
\end{array}
\eqno(\mathrm{A}.12)
$$
If one neglects the explicit $\epsilon$ dependence of
$\tilde\Sigma_{ij}(\mathbf{k},\epsilon)$, the integral
(A.9) vanishes.
On the other hand, from Eq.~(A.10) it follows that
the nonvanishing contribution into $L(\mathbf{p}, \mathbf{k}, \omega)$
is given by the odd in $\epsilon$
parts of the functions $C_{\pm}(\mathbf{p}, \mathbf{k}, \omega,\epsilon)$
and the even in $\epsilon$ parts of the functions
$D_{\pm}(\mathbf{p}, \mathbf{k}, \omega,\epsilon)$.
Since $\tilde\Sigma_{12}(\mathbf{p}, \mathbf{k})$
is an even function of $\epsilon$, and
$\tilde\Sigma_{11}(\mathbf{p}, \mathbf{k})$ contains both an even
$\tilde\Sigma_{11}^s(\mathbf{p}, \mathbf{k})$ and an odd
$\tilde\Sigma_{12}^a(\mathbf{p}, \mathbf{k})$ part,
the expressions for $C_{\pm}$ and $D_{\pm}$,
which provide for nonzero values of
$\Im \Pi(\mathbf{k}, \omega)$, can be cast into the form
$$
\begin{array}{c}
\displaystyle
\tilde C_{\pm}(\mathbf{p}, \mathbf{k}, \omega,\epsilon)=
\frac{1}
{E^2(k)-\left[\omega\pm E(\mathbf{k}- \mathbf{k})\right]^2}
\left\{
\tilde\Sigma_{11}^a(\mathbf{k}-\mathbf{p},\epsilon-\omega)
\right.
\\[12pt]
\displaystyle
\left.{}\times
\left[
E(k)+E(|\mathbf{k}-\mathbf{p}|)+\omega+
\tilde\Sigma_{11}^s(\mathbf{k},\epsilon)-\mu+\frac{\mathbf{k}^2}{2m}
\right] \right.
\\[12pt]
\displaystyle
\left.{}+\tilde\Sigma_{11}^a(\mathbf{k},\epsilon)
\left[
\tilde\Sigma_{11}^s(\mathbf{k}-\mathbf{p},\epsilon-\omega)
-\mu+\frac{\left( \mathbf{k}-\mathbf{p}\right)^2}{2m}
\right]\right\}\;;
\end{array}
\eqno(\mathrm{A}.13)
$$

$$
\begin{array}{c}
\displaystyle
\tilde D_{\pm}(\mathbf{p}, \mathbf{k}, \omega,\epsilon)=
\tilde\Sigma_{11}^s(\mathbf{k},\epsilon)+
\tilde\Sigma_{11}^s(\mathbf{k}-\mathbf{p},\epsilon-\omega)-2\mu+
\frac{\mathbf{k}^2}{2m}+\frac{(\mathbf{k}-\mathbf{p})^2}{2m}-\omega
\\[12pt]
\displaystyle
{}+
\left[\omega\pm
E(|\mathbf{k}-\mathbf{p}|)\right]
\left\{
\tilde\Sigma_{11}^s(\mathbf{k}-\mathbf{p},\epsilon-\omega)
\left[
E(k)+E(|\mathbf{k}-\mathbf{p}|)+\omega+
\tilde\Sigma_{11}^s(\mathbf{k},\epsilon)-\mu+\frac{\mathbf{k}^2}{2m}
\right] \right.
\\[12pt]
\displaystyle
\left.{}
+
\tilde\Sigma_{11}^a(\mathbf{k},\epsilon)
\tilde\Sigma_{11}^a(\mathbf{k}-\mathbf{p},\epsilon-\omega)+
\tilde\Sigma_{12}(\mathbf{k},\epsilon)
\tilde\Sigma_{12}(\mathbf{k}-\mathbf{p},\epsilon-\omega)
\right\}\;.
\end{array}
\eqno(\mathrm{A}.14)
$$
At the same time, a nonzero value of $\Im\,\Pi(\mathbf{p},\omega)$
implies that the retarded renormalized potential (\ref{22})
becomes complex:
$$
\begin{array}{c}
\displaystyle
\Re\,\tilde V(\mathbf{p}, \omega)=
V(p)\frac{1-V(\mathbf{p})\Re\,\Pi(\mathbf{p}, \omega)}{
\left[1-V(\mathbf{p})\Re\,\Pi(\mathbf{p}, \omega)\right]^2+
\left[V(\mathbf{p})\Im \Pi(\mathbf{p}, \omega)\right]^2}\;;
\end{array}
\eqno(\mathrm{A}.15)
$$

$$
\begin{array}{c}
\displaystyle
\Im\,\tilde V(\mathbf{p}, \omega)=
\frac{-V^2(\mathbf{p})\Im\,\Pi(\mathbf{p}, \omega)}{
\left[1-V(\mathbf{p})\Re\,\Pi(\mathbf{p}, \omega)\right]^2+
\left[V(\mathbf{p})\Im\,\Pi(\mathbf{p}, \omega)\right]^2}\;.
\end{array}
\eqno(\mathrm{A}.16)
$$
Therefore, the functions
$\tilde \Psi_{11}(\mathbf{p},\omega)$ and $\tilde \Psi_{12}(\mathbf{p},\omega)$
become complex, too, and so does the quasiparticle spectrum
$\tilde E(p)=
E(p)+i\gamma(\mathbf{p})$, where $\gamma$
is the decrement of damping.

The dynamical structure factor
$$ S(\mathbf{p}, \omega)=-\frac{1}{\pi n}
\Im\,\tilde\Pi(\mathbf{p}, \omega)\;,
$$
which is measured in neutron scattering experiments and
is usually determined by the imaginary part of
the density-density correlation function $\tilde\Pi(\mathbf{p}, \omega)$
(see, for instance, \cite{?}), can be also obtained as the
imaginary part of the ``screened'' interaction
$$ S(\mathbf{p}, \omega)=-\frac{1}{\pi V(q)}
\Im\,\tilde V(\mathbf{p}, \omega)\;.
$$
On the ``mass shell'', the form factor $S(p)\equiv S(p, E(p))$
has a maximum in the same region of $p$ where the potential
$\tilde V(\mathbf{p}, \omega)$ and correspondingly
the quasiparticle spectrum $E(p)$ have a minimum, in accordance
with the general conection beetween $E(p)$ and $S(p)$ (see
\cite{Feynman}--\cite{Pitaevskiy2}).

\newpage

\begin{center}
\includegraphics[0,0][349,309]{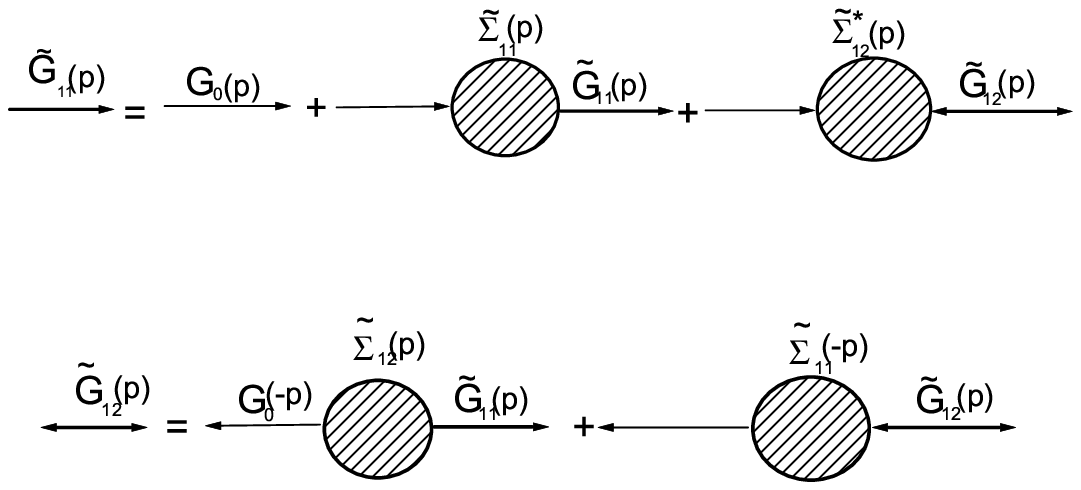}
\end{center}

Fig.~1.
The diagram representation of the Dyson--Belyaev
equations for the normal $\tilde G_{11}$ and anomalous $\tilde G_{12}$
Green functions of bosons in the superfluid Bose system.

\newpage

\begin{center}
\includegraphics[0,0][339,278]{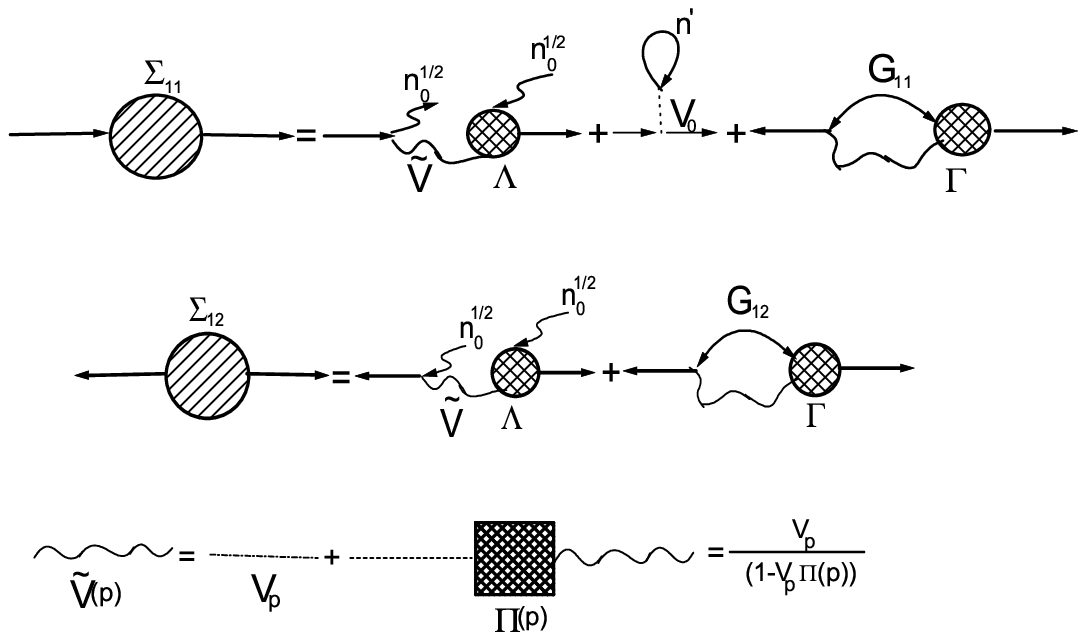}
\end{center}

Fig.~2.
The diagram representation of the nonlinear integral
equations for the normal $\tilde \Sigma_{11}$ and anomalous
$\tilde \Sigma_{12}$ self-energy parts in the Bose liquid
with account for terms of first order in the small density of
the single-particle BEC $(n_0\ll n)$ and of the equation
for the renormalized (``screened'') potential of pair
interaction $\tilde V$.

\newpage

\begin{center}
\includegraphics[0,0][336,258]{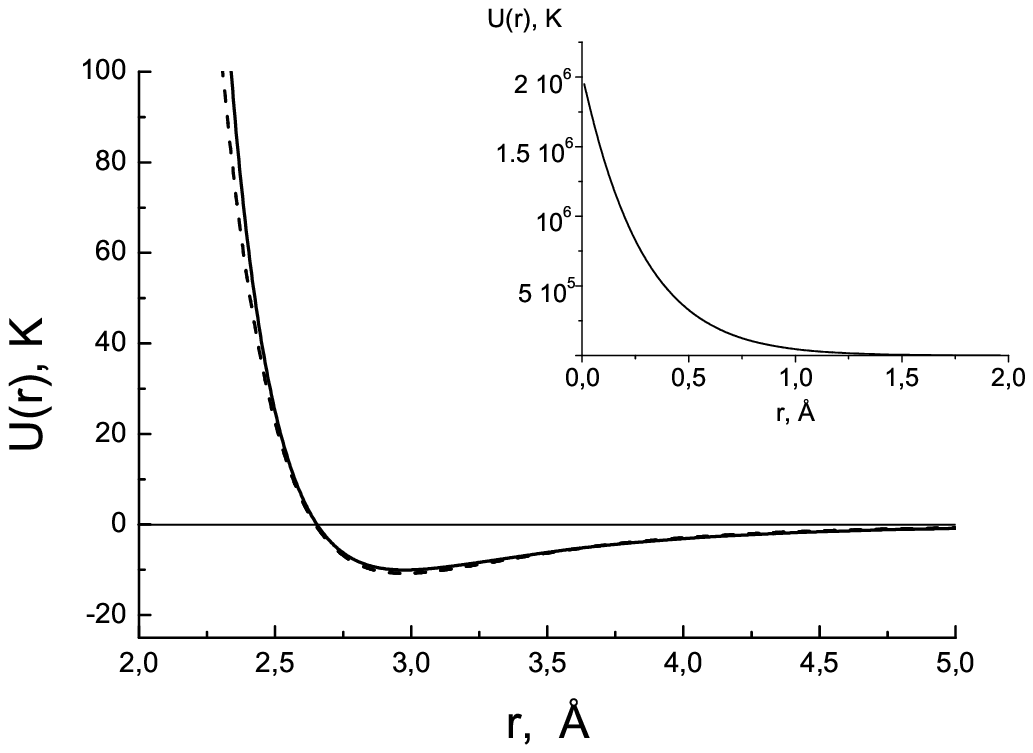}
\end{center}

Fig.~3.
The radial dependences of the regularized Aziz potential (\ref{Aziz})
(solid curve) and the Lennard-Jones potential (\ref{37}) (dashed
curve). Inset, the Aziz potential at small distances, down to $r=0$.

\newpage

\begin{center}
\includegraphics[0,0][252,341]{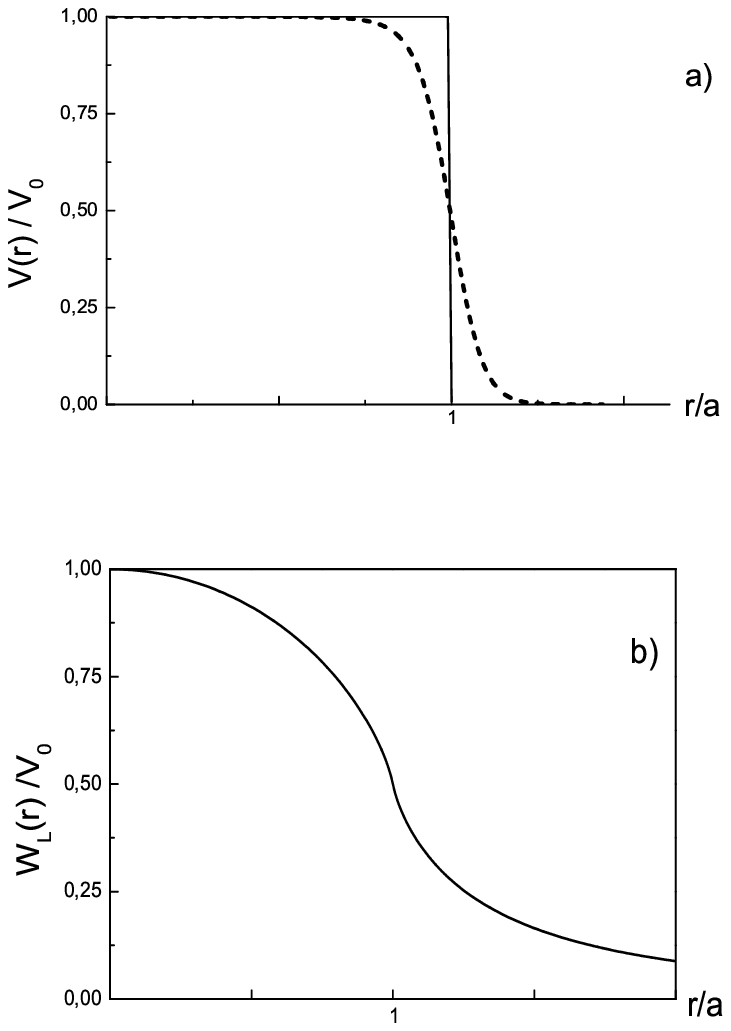}
\end{center}

Fig.~4.
The finite repulsion potential in the model of
``soft spheres'' in the form of (a) a Fermi ``step''
(\ref{39}) (solid curve corresponds to $b=0$, dashed
curve---to $b=0.5a$), and (b) a Lindhardt function (\ref{41})
in real space.

\newpage

\begin{center}
\includegraphics[0,0][336,255]{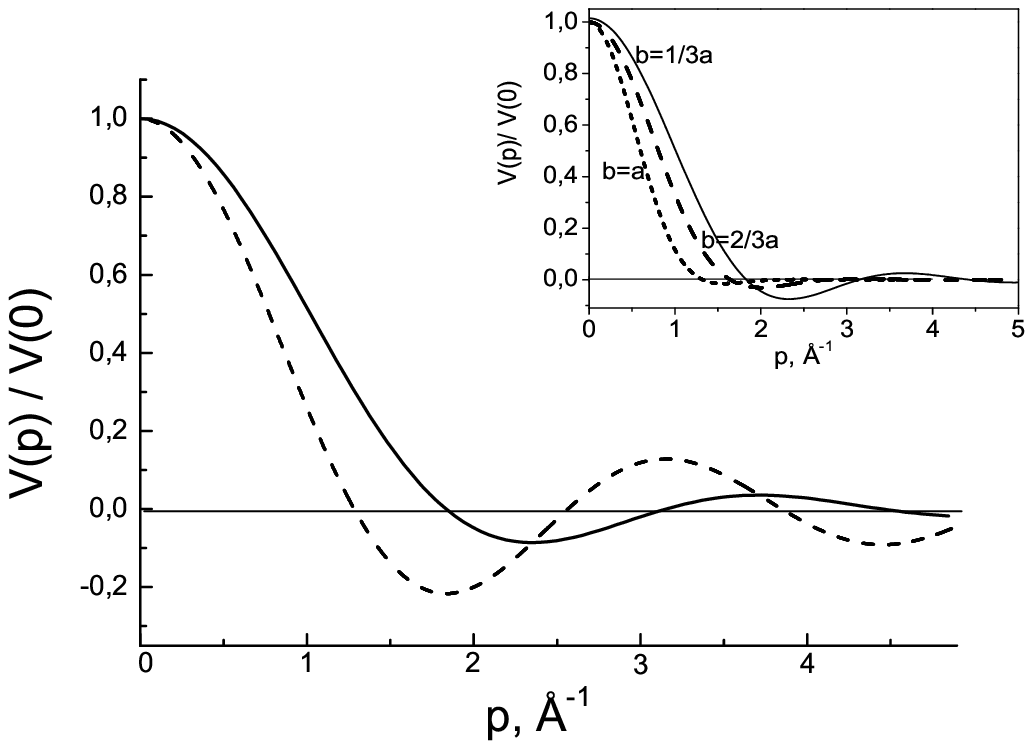}
\end{center}

Fig.~5.
The momentum dependences of the Fourier component (\ref{40}) of the Fermi-
type potential (\ref{39}) at $b=0$ or the Lindhardt-type
potential (\ref{41}) (solid curve) and of the
potential of the ``hard spheres'' model \cite{28}
$V(p)=V_0\sin(pa)/pa$ (dashed curve). Inset is the Fourier component
for the Fermi type potential with $b\ne 0$.

\newpage

\begin{center}
\includegraphics[0,0][298,235]{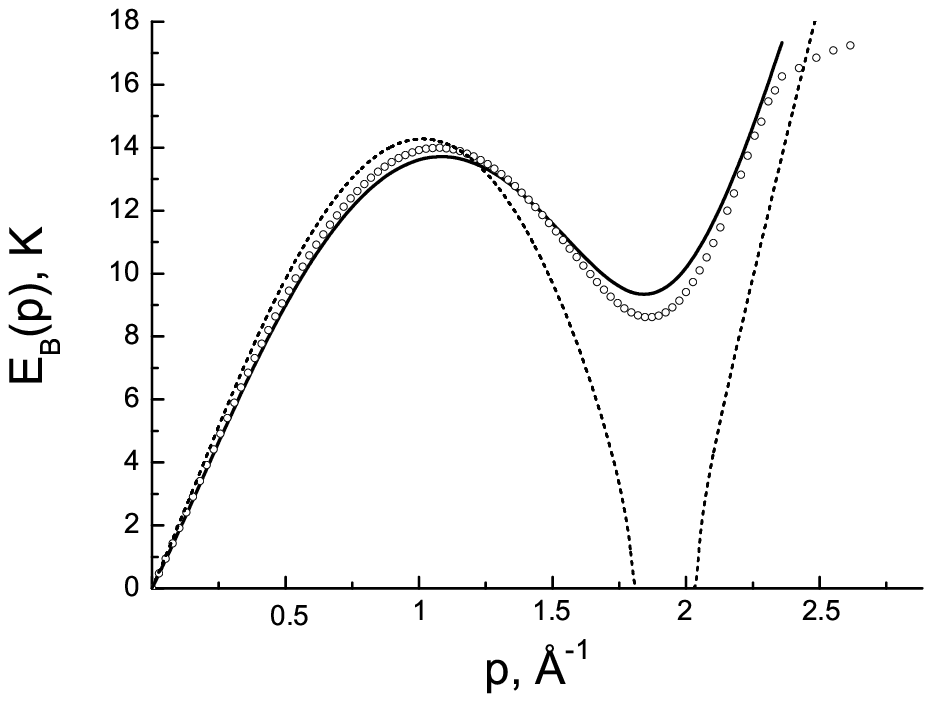}
\end{center}

Fig.~6.
The Bogolyubov quasiparticle spectrum (\ref{42}) with an oscillating
potential (\ref{40}) (solid curve), maximally close to the experimental spectrum
\cite{30}--\cite{32} (dotted curve) at $V_0/(4\pi a^3)=169$~K and $a=2.44$~\AA.
The dashed curve corresponds
to the unstable spectrum (\ref{42}) at $V_0/(4\pi a^3)=217$~K,
characterized by negative values
$E^2_B(p)<0$ in some momentum interval.

\newpage

\begin{center}
\includegraphics[0,0][309,233]{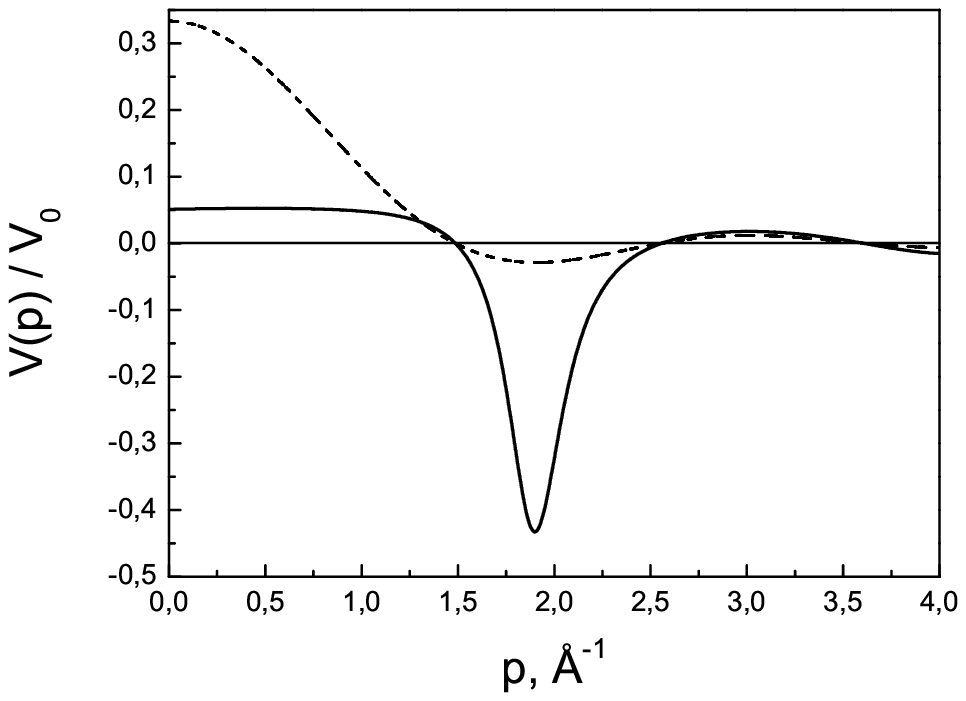}
\end{center}

Fig.~7.
The momentum dependence of the renormalized (``screened'') potential
(\ref{22}) (solid curve) with account for the momentum dependence both 
of the Fourier component (\ref{40}) (dashed curve) and of the polarization operator
$\Pi(p,E(p))$ on the ``mass shell'' (see Fig.~8).

\newpage

\begin{center}
\includegraphics[0,0][301,236]{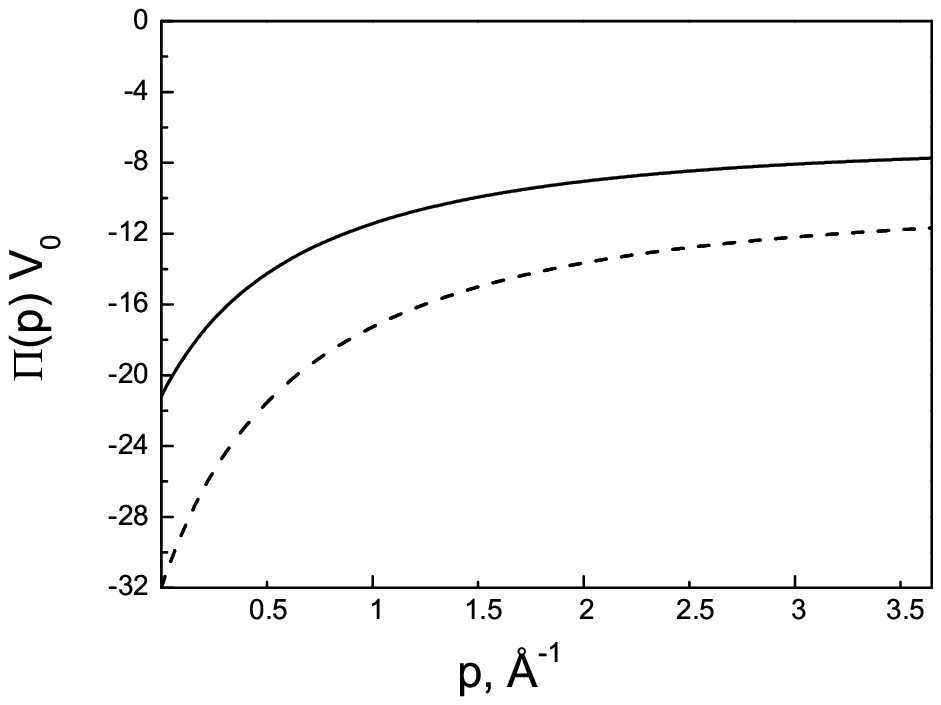}
\end{center}

Fig.~8.
The momentum dependences of the real part of
the boson polarization operator on the ``mass shell'',
$\Pi(p)\equiv \Re\,\Pi(\mathbf{p}, E(p))$, multiplied by $V_0$,
obtained from self-consistent computations at $\Gamma=1$
(solid curve), and of $V_0\Pi(p)\Gamma$ at $\Gamma=1.5$
(dashed curve).

\newpage

\begin{center}
\includegraphics[0,0][295,228]{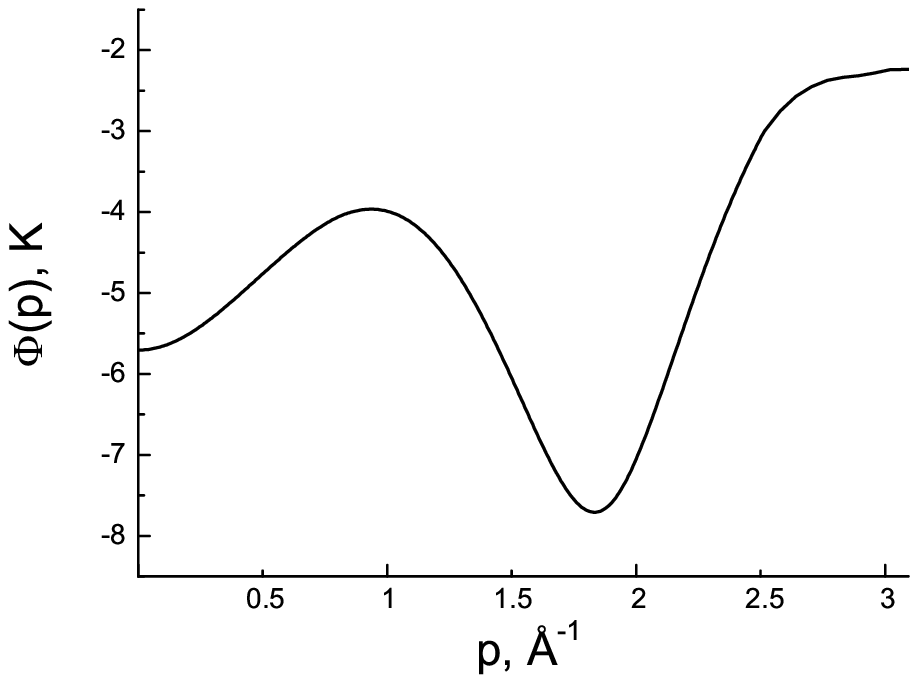}

(a)

\includegraphics[0,0][296,233]{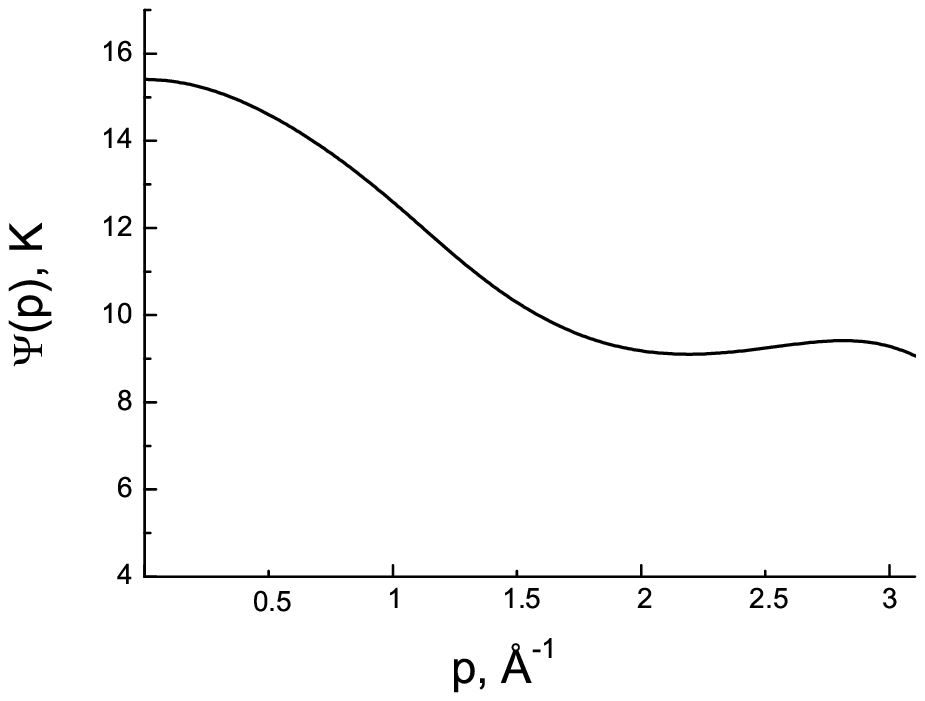}

(b)

\includegraphics[0,0][292,231]{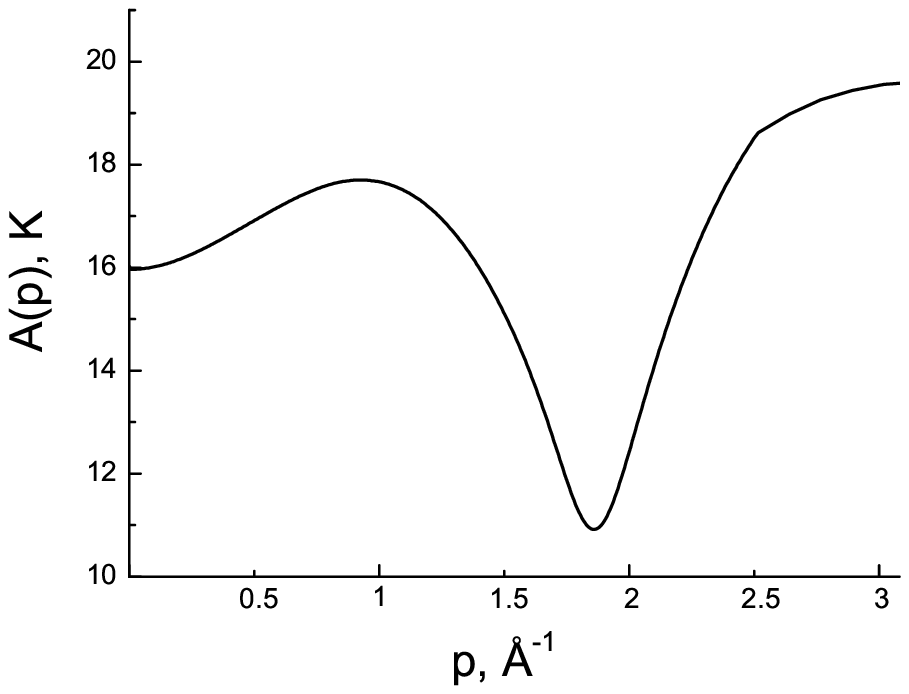}

(c)

\end{center}

Fig.~9.
The momentum dependences of the functions
$\Phi(p)$ (a),  $\Psi(p)$ (b),  $A_0(p)$ (c), obtained from
self-consistent computations at the value of the single
fitting parameter $V_0/(4\pi a^3)=1552$~K for $a=2.44$~\AA.

\newpage

\begin{center}
\includegraphics[0,0][296,231]{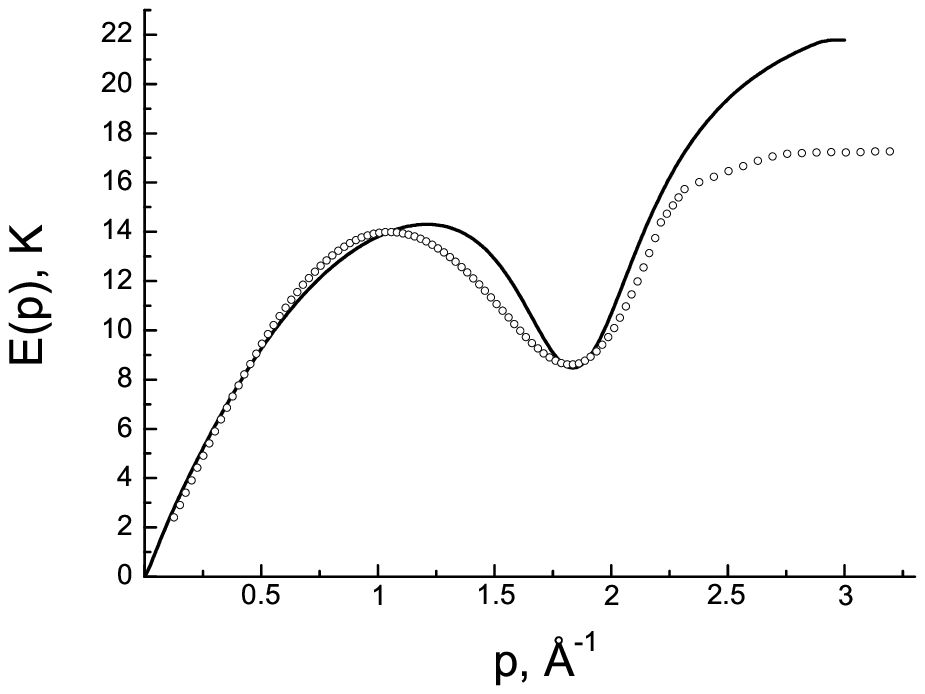}
\end{center}

Fig.~10.
The theoretical quasiparticle spectrum $E(p)$ obtained from the
self-consistent computations in the model of ``soft spheres'' for
a quantum Bose liquid vith a suppressed BEC in the approximation of
constant vertex $\Lambda(p)=\Gamma(p,0)=1.5$ for all $p$ (solid curve)
and the empirical spectrum of elementary exitations in the SF
Bose liquid $^4$He \cite{30}-\cite{32} (circles).

\newpage

\begin{center}
\includegraphics[0,0][302,227]{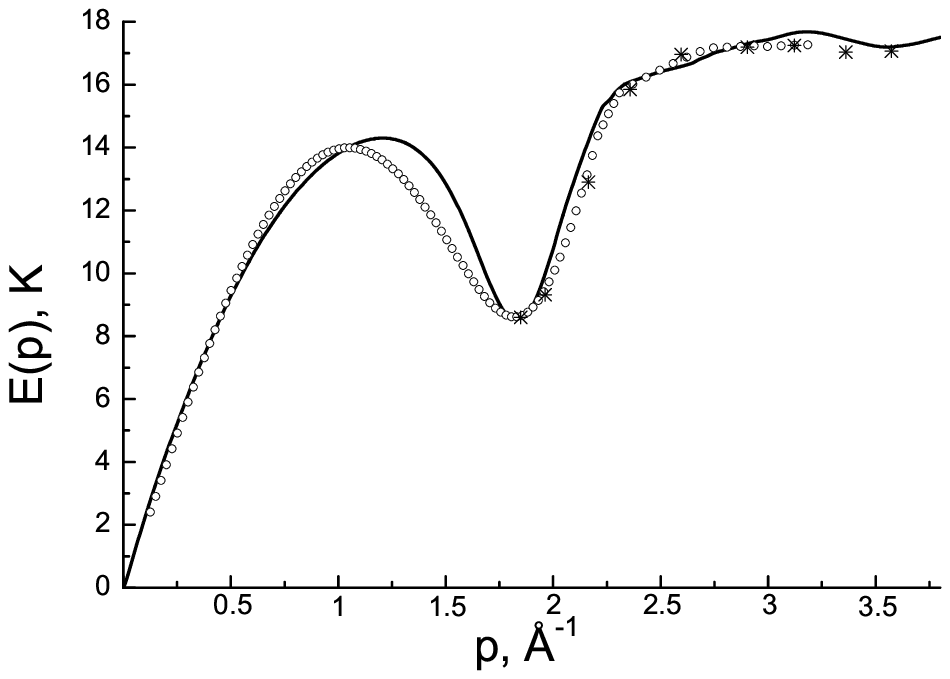}
\end{center}

Fig.~11.
The theoretical quasiparticle spectrum $E(p)$ calculated with the
vertex being a decreasing function falling off from $\Gamma=1.5$
to $\Gamma=1.1$ in the region 2.1~\AA~$<p<$~3.8~\AA (solid curve)
and the empirical spectra \cite{30}-\cite{32} (circles) and \cite{Pearce}
(asterisks).

\end{document}